# Neutron Spectroscopic Factors from Transfer Reactions

*Jenny Lee, M.B. Tsang, and W. G. Lynch*

*National Superconducting Cyclotron Laboratory and Department of Physics and Astronomy, Michigan State University, East Lansing, MI 48824*

**Abstract**

The present paper examines past measurements of the angular distributions for (d,p) and (p,d) reactions on targets with Z=3-24 leading to the ground states. The procedure is prescribed for extracting a conventional set of spectroscopic factors. Most of these spectroscopic factors agree well with large-basis shell model predictions. In all, the ground state neutron spectroscopic factors for 80 nuclei have been obtained. The consistency of the method is evaluated by comparing spectroscopic factors obtained separately in (p,d) and (d,p) reactions. The values correlate with Endt's compilation when available, but the current method is more general and the spectroscopic factor values obtained are more consistent with each other.

## I. Introduction

Spectroscopic factors describe the single particle structure of nuclei in the shell model. It is defined as the overlap between the initial and final state in the reaction channels [1-5]. In the past four decades, single nucleon transfer reactions such as (d,p) or (p,d) reactions have been used extensively to extract the spectroscopic information of the single nucleon orbits [1-6]. Specifically, these measurements allow the extraction of the spectroscopic factors by taking the ratios of the experimental cross-sections to the predicted cross-sections from a reaction model. The most common model used is the Distorted Wave Born Approximation (DWBA) theory [3-5]. For (p,d) and (d,p) transfer reactions involving deuteron, the effects from deuteron break up can be significant at high energy and the correction is generally taken into account using the Johnson-Soper adiabatic approximation [7] to construct the deuteron potential. As this approach is not strictly DWBA, we will call it the adiabatic three-body model.

It is not unusual to find published spectroscopic factors for a particular nucleus that fluctuate by factors of 2-3 [8]. On the other hand, there are published spectroscopic factors, which are the same within the quoted uncertainties even though the data used to extract them do not agree with each other to factors of 2-3. Some of the difficulties in the past extractions of spectroscopic factors have been associated with ambiguities in the parameterizations used in the reaction models, different normalizations, and different assumptions used in the analysis [8, 9]. To allow comparisons of the experimental spectroscopic factors with theoretical predictions over a broad range of nuclei, it is important that a systematic and consistent approach be developed to analyze transfer reactions data. Furthermore, as transfer reaction continues to be one of the important techniques to elucidate the structure of exotic nuclei, it is important to develop analysis techniques that can be extended to nuclei far from stability.

In a large-scale surrey of 80 nuclei studied via the transfer (p,d) and (d,p) reactions [10], the ground state spectroscopic factors have been deduced using an adiabatic three-body model as described above. Most of the extracted SF values agree with the predicted SFs from large-basis shell-model (LB-SM) calculations within the experimental and theoretical uncertainties. These spectroscopic factors obtained over a wide range of nuclei provide important benchmarks against which more advanced methods for studying the nucleon transfer reaction mechanisms can be compared [11-12]. For such study, it is important to know which sets of data should be included in future analysis and where the gap of knowledge may lie.

The data published in ref. [10] were obtained from transfer reaction data collected in the past 40 years. Of the 423 reactions studied, only 230 were used to extract the SF values. One purpose of this paper is to set forward the criteria used in our data evaluation and the quality control measures that we applied to select the 230 reactions out of the larger set of 423 reactions that have been measured by many research groups. In addition, this paper presents a procedure whereby a "conventional" set of spectroscopic factors can be obtained from (p,d) and (d,p) transfer reactions. The set of spectroscopic factors obtained agree well with the modern day shell model predictions and can be viewed as benchmarks for other analysis with different input or analysis criterira.

This paper is organized as follows. We begin in Section II with a brief description of the input parameters used in three-body adiabatic reaction model. This is important because spectroscopic factors are extracted by dividing the measured differential cross-sections with the theoretical cross-sections predicted by a reaction model. We then explain in Section III how the data have been compiled and the uncertainties introduced in the process. We explain in Section IV the procedure for extracting the SFs. Problems with consistencies between measurements are discussed in Section V-VII. As the pickup (p,d) reaction is the inverse of the stripping (d,p) reaction, the SFs obtained separately by the (p,d) and (d,p) reactions should be the same within experimental uncertainties. We use this fact in Section VIII to assess the consistency of our method and to assign uncertainties to the extracted SFs. Section IX compares some of our SF values with those compiled by Endt [9]. Due to recent interest in the neutron spectroscopic factor of $^{15}$C, Section X discusses the challenges and problems of the reaction, $^{14}$C(d,p) $^{15}$C. Recently, it has been observed in nucleon-knockout reactions that spectroscopic factors are suppressed with respect to the modern LB-SM values with increasing nucleon separation energy. Section XI discussed whether there is evidence for such trend in the transfer reaction data we analyzed. Section XII summarizes our findings.

## II. Reaction model

In the following, we have adopted the conventional analysis widely used in the literature on neutron-transfer reactions and have tried to apply it consistently over the range of reactions studied, with minimal assumptions. In the present work, we follow the algorithm developed in ref. [8] and use a modified version of the code TWOFNR [13] to perform the transfer reaction model calculations using the same input parameters labeled as CH in ref. [8]. The code TWOFNR is chosen mainly for convenience as it contains all the input options discussed below. Other widely used reaction model codes, DWUCK5 and FRESCO yield nearly identical predictions with the same input parameters [11, 12, 14].

The transfer cross-sections are calculated within the Johnson-Soper (JS) adiabatic approximation [7] to the neutron, proton, and target three-body system using the phenomenological nucleon nucleus optical model potentials [15]. This calculation includes the effects of breakup of the deuteron in the field of the target. The valence neutron binding potential is Woods-Saxon in shape with a fixed radius parameter of 1.25 fm and a diffuseness parameter of 0.65 fm [8]. The depth of the potential is normalized to the experimental binding energy. All calculations make the local energy approximation (LEA) for finite range effects [16] using the Zero-range strength ($D_o^2$=15006.25 MeV$^2$ fm$^3$) and range ($\beta$=0.7457 fm) parameters of the Reid soft-core $^3S_1$-$^3D_1$ neutron-proton interaction [17]. Nonlocality corrections with range parameters of 0.85 fm and 0.54 fm are included in the proton and deuteron channels, respectively [18]. The same set of input parameters is used throughout in the present work to extract the spectroscopic factors. Since the input parameters adopted in this work have been widely used in the literature, we label our SF values as SF(conv) in our figures and in Table III, to distinguish them from other SF values obtained when different input parameters are used.

## III. Compilation and digitization of angular distribution data

For the present work, we mainly focus on the transfer reaction A(d,p)B and its inverse reaction B(p,d)A where the nucleus A is considered to be composed of the core B plus the valence neutron n. Table 1 contains 423 reactions that we have examined. For clarity, we include shorthand literature references [19-240] in the table.

Nearly all the angular distributions listed in Table I have been digitized from the published figures. The few exceptions are those found in the Nuclear Science References (NSR) database of the National Nuclear Data Center (NNDC) [241]. The data from NSR are in tabulated form and the sources of these data came from the Former Soviet Union or Japan whose journals are not widely available in the United States. These non-US and non-European data complement our search in the Physical Review, Physical Review Letters, Nuclear Physics and occasionally in Physics Letters. While we have made an effort to find nearly all the relevant experiments that published the absolute differential cross-sections, we may have missed some reactions especially if the incident energy is below 10 MeV and above 70 MeV. Except when noted, the table does not include reactions with cross-sections published in arbitrary units. The data and calculations are posted in a website [242]. Eventually, we hope all the digitized data used in this work will be adopted by the NSR.

By checking some of the data carefully and sometimes repeating the digitization several times, we estimate the uncertainties introduced by the digitization process to be less than 10%. For illustration, we use the data for the reaction $^{14}N(d,p)^{15}N$ at $E_d$=12 MeV [21, 82]. This set of data was first published in tabulated form in ref. [21]. The tabulated data are plotted as closed points in Figure 1. Later the authors in ref. [82] plotted the data in a figure. We digitize the data in ref. [82] and compare our digitized data (open points) with the tabulated data (closed points) in Figure 1. We see a difference of less than 10% between the two sets of data. Of course, the digitization errors also depend on the actual size of the graphs available in the original literature. As described later, generally, errors introduced by digitization are small compared to the uncertainties in the absolute cross-section measurements.

**IV. Extraction of spectroscopic factors**

For nearly all the nuclei we study, we use the ground state $\ell$ values determined from the angular distributions and the $\mathbf{j}^{\pi}$ values of the valence neutron ground states found in the isotope tables [243]. In general, the experimental angular distributions at backward angles are more sensitive to the effects of inelastic couplings and other higher-order effects and are not well reproduced by most reaction models. Furthermore, discrepancies between the shapes from calculations and experiment are much worse at the cross-section minimum. Thus, we follow the procedures developed in ref. [8] and others that the spectroscopic factor is extracted by fitting the reaction model predictions to the angular distribution data at the first peak, with emphasis on the maximum. The accuracy in absolute cross-section measurements near the peak is most important. When possible, we take the mean of as many points near the maximum as we can to extract the spectroscopic factors. We will use the angular distributions of $^{14}N(d,p)^{15}N$ shown in Fig 1 to illustrate the procedure we adopt to extract the spectroscopic factors.

In Figure 1, the first three data points with $\theta_{cm}<25°$ have been used to determine the ratios of the measured and calculated differential cross-sections. The mean of these three ratios is adopted as the spectroscopic factor. For example, for the two sets of data plotted in Figure 1, the spectroscopic factors are 1.2 and 1.1 for digitized [82] data and tabulated data [21] respectively. The difference in the spectroscopic factors represents the uncertainties introduced by digitizations. The theoretical angular distributions, obtained from TWOFNR, multiplied by the spectroscopic factor 1.1, are plotted as solid curve in the figure.

In cases when a "first peak" is not obvious or that the angular distributions of the forward angles are nearly flat, e.g. in the reaction of $^{44}Ca(p,d)^{43}Ca$ at $E_p$=40 MeV [174] as shown in Figure 2, we find that fitting the shoulder gives more consistent results. In general, the agreement of the shape of the angular distributions of the first peak or shoulder to reaction calculations gives some indication as to the quality of the spectroscopic information that can be extracted by comparing the transfer model to the data. When appropriate, the number of data points from a given measurement that lie in the region where data can be described well by the transfer model is taken to compute the relative weights of SF's extracted from different measurements that could combine together to get the mean spectroscopic factors presented here.

**V. Evaluation of the angular distribution measurements**

Even though most papers state the uncertainties of their cross-section measurements to be 10-20%, the actual disagreements between experiments are often larger than the quoted uncertainties. An example is illustrated in the reactions $^{11}B(d,p)^{12}B$ reactions. From the conventional literature, we find two measurements at deuteron incident energy of 11.8 MeV [45] and 12 MeV [21]. Since the incident deuteron energy is nearly the same, one would expect the angular distributions from the two data sets plotted in Figure 3 to be the same within experimental error. Ref. [21] (open circles) stated that the accuracy of the absolute cross-section measurements is 15% while ref. [45] (closed circles) quoted an error of 6%, which is smaller than the closed symbols in Fig 3. Not only do the cross-sections differ sometimes by a factor of two, the shapes of the distributions (especially the first peak) are not even the same. In this case, the shape of the angular distributions in ref. [45] agrees with the calculation (solid curve) better than that measured in ref. [21]. Fortunately for this reaction, we are able to find another measurement in the NNDC database [46] (open diamonds). Near the peak at forward angles, this latter angular distribution agrees with ref. [45] and so we disregard the measurements of ref [21]. Data in ref. [45] were measured nearly 40 years later than data in ref. [21] and one might be tempted to attribute the difference to the availability of better beam quality and detection systems for the measurements. However, when another reaction, $^{12}C(d,p)^{13}C$ at $E_d$=11.8 MeV from ref. [45] (closed circles) is compared to three other published angular distributions in Figure 4 at $E_d$=11.8 MeV (closed diamonds) [30], 12 MeV (open circles) [21], 12 MeV (open diamonds) [59], the cross-sections in the first peak measured in ref. [45] is consistently low. No uncertainties in the measurements are given in ref. [30] and ref. [59] but it is clear that data in ref. [45] do not agree with the other measurements, especially in the most forward angle region. Thus we disregard the SF derived from ref. [45] in our compilation of $^{12}C(d,p)^{13}C$ reactions.

Cross comparisons of angular distributions sometimes help to establish common systematic problems when one set of measurements was performed by the same group with the same set up. An example is illustrated in the $^{40}Ca(d,p)^{41}Ca$ reactions in ref. [181] where the ground state angular distributions of $^{41}Ca$ at $E_d$=7, 8, 9, 10, 11 and 12 MeV have been measured. Figure 5 shows the extracted spectroscopic factors (labeled as SF(conv)) as a function of incident deuteron energy for all the $^{40}Ca(d,p)^{41}Ca$ reactions. For clarity in presentation, no error bars are plotted. Except for the point at $E_d$=7 and 12 MeV, the extracted spectroscopic factors from ref. [181] (open circles) are consistently larger than the spectroscopic factors extracted from different experiments with the same reaction at the same energy. Detailed comparisons of the angular distribution data show essentially the same effect, that the differential cross-sections measured in ref. [181] are systematically higher than the other measurements [30, 178, 182-191]. Clearly, there must be some problems in the determination of the absolute cross-sections in ref. [181]. Since it is not possible to find the cause after so many years, in our review of the data, we disregard the spectroscopic factor values determined in ref. [181].

Similarly we disregard the data in ref. [29] for the $^{9}Be(d,p)^{10}Be$ reaction as most of the data in ref. [29] are low when compared to the available data from other measurements. There are other examples. All the SF values not used are listed in Column 5 of Table I. In general, a brief comment follows in the last column of Table I if the data set is considered to be problematic.

The disagreements between data sets generally exceed the quoted uncertainties of the experimenters. Indeed, we have found that the most important aspect of quality control of the data is to have as many independent measurements as possible. Comparisons of different measurements help to identify problematic measurements. The large number of measurements compiled in Table I have helped to improve the quality of the spectroscopic factors extracted in the present work.

**VI. Transfer reactions at high and low energy**

When Q-value and the transverse and angular momentum transferred are not well-matched as in the transfer reactions induced by very low or high (> 50 MeV) beam energy, the shape comparisons are also poor. Figure 6 shows the angular distributions of the protons emitted from the $^{40}Ca(d,p)^{41}Ca$ (g.s) reaction from $E_d$=7.2 to 56 MeV. Only one angular distribution is presented at each energy. The agreement between data and prediction for the first peak improves with increasing energy. At very low incident or excitation energy, the shapes of the measurements and the calculated transfer cross-sections do not agree. This phenomenon is also seen in other reactions. The spectroscopic factors as a function of incident energy are shown in Figure 5. The increase of spectroscopic factors at $E_d$<10 MeV has been observed before [8, 21] and has been attributed to the resonance structures in the elastic scattering of the deuterons [244]. As explained in the last section, the open points based on the data from ref. [181] are discarded. Between 10 to 56 MeV, the mean spectroscopic factor, 1.01 ± 0.06 as shown by the solid line in Figure 5, is independent of incident energy within experimental errors.

In reactions which have large negative Q values such as $^{12}C(p,d)^{11}C$ (Q = –16.5 MeV), the center of mass energy available in the exit channel is very small even at ~20 MeV proton incident energy [38]. The validity of the calculated angular distribution is questionable at these energies and we discard these data. For other reactions measured at low incident energy (<10 MeV), the data could be dominated by compound nucleus emissions, or resonances in the low energy elastic scattering [244]. When possible, we exclude spectroscopic factors obtained with incident beam energy less than 10 MeV. These "excluded" spectroscopic factors, not included in computing the mean values of the spectroscopic factors, are listed in Column 5 of Table I.

Even though we exclude data with incident energy lower than 10 MeV from the calculation of the mean SF, these low energy data are still valuable. In cases where very few (sometimes only one) measurements with incident energy greater than 10 MeV are available, they provide checks for consistency of the measurements. Examples are $^{49}Ti(p,d)^{48}Ti$ and $^{48}Ti(d,p)^{49}Ti$ reactions [146, 211, 217, 218]. In the $^{43}Ca(d,p)^{44}Ca$ reaction, only 8.5 MeV data [198] are available. Similarly, only 7.5 MeV data for $^{50}V(d,p)^{51}V$ reaction [220] and 7.83 MeV data for $^{23}Na(d,p)$ $^{24}Na$ reaction [110] are found. We adopt these results nonetheless.

At high energy, momentum transfer and angular momentum transfer are mismatched so conditions may not be optimized to extract reliable spectroscopic factors for the ground state valence neutrons. Furthermore, the global nucleon-nucleus potentials (CH89) [15] are fitted only to 65 MeV for protons and 26 MeV for neutrons. Thus, we do not include high energy reactions in this work. In examining data over a wide range of d or p incident energies, we find that the optimum beam energies for studying transfer reactions lie between 10-20 MeV per nucleon.

## VII. Nuclei with small spectroscopic factors

For the $^{50}Cr(p,d)^{49}Cr$ reactions, there are two measurements at beam energy of 17.5 and 55 MeV [223, 224]. In each case, the predicted and measured angular distributions are different as shown in Figure 7 with closed circles for 17.5 MeV [223] data and open circles for 55 MeV data [224]. From the magnitude of the measured cross-sections, a spectroscopic factor value of 0.11 is derived. The extracted spectroscopic factor is very low especially for an even-even nucleus. It is reasonable to speculate that there is considerable configuration mixing of the valance nucleus. When very low SF values (compared to values predicted by the Independent Particle Model [3-5]) are obtained, data quality may be poor and the predicted shape of the angular distributions may not agree well with that of the data. This may indicate that one step transfer amplitudes are not dorminant and comparison of data to such calculations may be unreliable. Other examples are $^{20}$F, $^{21}$Ne, $^{22}$Ne, $^{24}$Mg, $^{35}$Cl, $^{45}$Sc, $^{47}$Ti, $^{48}$Ti, $^{50}$Cr, and $^{51}$V nuclei.

In the case of $^{46}$Ti(d,p)$^{47}$Ti reactions [214, 215], both measurements at $E_d$=7 and 10 MeV are very different from the predicted cross-sections and disagree with each other in shape and absolute cross-sections. We did not extract spectroscopic factors for this nucleus.

**VIII. Comparison of Spectroscopic factors obtained from (p,d) and (d,p) reactions**

The neutron pickup (p,d) and neutron stripping (d,p) reactions are inverse reactions, both of which connect the ground states of the target and projectile nuclei. They should yield the same values of the spectroscopic factors. From Table I, we select the nuclei, which have been studied reasonably well by both neutron pick-up and stripping reactions from and to the ground state. The averaged SF values are listed in the 2$^{nd}$ and 4$^{th}$ column of Table II. The numbers of measurements contributing to the averages are listed next to the mean values in the 3$^{rd}$ and 5$^{th}$ column.

There are strong correlations between the spectroscopic factors determined from the (p,d) and (d,p) reactions as shown in Figure 8. The solid line indicates perfect agreement. As these are independent measurements determined from similar procedure outlined above, we use the scatter of the data points to determine the errors. Assuming the uncertainties of each measurement are the same, and requiring the chi-square per degree of freedom to be unity, we can extract a random uncertainty of 20% for a given measurement. The obtained uncertainty of 20% is consistent with comparisons with analysis on systems that have large number of measurements such as $^{12}$C(d,p)$^{13}$C, $^{16}$O(p,d)$^{15}$O, $^{16}$O(d,p)$^{17}$O, $^{40}$Ca(d,p)$^{41}$Ca and other reactions. Examinations of large number of measurements in Table I suggest that the uncertainties in the extraction of the spectroscopic factors are largely limited by the disagreement between measurements. In Table II and Figure 8, we have excluded measurements for $^{7}$Li, $^{34}$S and $^{10}$Be nuclei due to large errors associated with either the (p,d) or (d,p) measurements. If we include these three measurements, the estimated uncertainty in a given measurement increases to 28%

Finally, we can compute the spectroscopic factor values and the associated uncertainties listed in Table III. The SF values are obtained from the weighted average of independent measurements from both the (p,d) and (d,p) reactions listed in Table I from which the low energy (<10 MeV) and outlier data (nominally marked with asterisks) are excluded. For values determined by only one measurement when no other independent measurement is available for consistency checks, an associated uncertainty of 28% is assigned. For values determined by more than one measurement (N), we take into account the distribution of the SF's around the mean. Figure 9 illustrates this procedure, the open stars represent the spectroscopic factors extracted from the good measurements of the calcium isotopes. However, the spread of the data are more than 20% for the $^{44}$Ca and $^{48}$Ca nuclei even though three "good" measurements are found for each of these nuclei. For these nuclei, it is more realistic to assign the uncertainty using the standard deviations of the mean of the data points. The associated uncertainties listed in Table III are determined by adopting either the standard deviations of the mean or the computed uncertainties determined by 20%/$\sqrt{N}$, whichever is larger. For comparison, the mean SF values with the associated uncertainties are plotted as the solid stars with error bars in Figure 9.

**IX. Comparison with Endt's "best values"**

In 1977, Endt compiled a list of the "best" spectroscopic factor values for the sd-shell nuclei [9]. For the neutron spectroscopic factors, Endt compiled the published spectroscopic factors from (d,t), (p,d), ($^3$He,$\alpha$) and (d,p) reactions. An uncertainty of 25% is assigned to the values. (When only the (p,d) and (d,p) reactions are studied, Endt assigned a 50% uncertainties.) Endt's best values are listed in Table III. Figure 10 compares the spectroscopic factors determined by Endt and the present work (SF(conv)). There are strong correlations between the two procedures even though the values scatter around the dashed line, which indicates perfect agreement. From the consistency check with (p,d) and (d,p) reactions, we expect that our values should have smaller random uncertainties because a systematic approach is used to extract the SF values directly from the measured angular distributions while Endt's compilation depended on the analysis by different authors and relied on the communication with the authors concerning the normalizations of the spectroscopic factors. We also have the advantage that many more measurements are included in Table I than those that were available for Endt's compilations.

## X. $^{14}C(d,p)^{15}C$ reactions

The $^{14}C(d,p)^{15}C$ reaction is an important reaction because $^{15}C$ has a loosely bound halo neutron. It is used to provide cross-comparisons between the spectroscopic factors obtained from one-nucleon knock-out and transfer reactions [245]. In addition, this reaction is a good candidate to extract spectroscopic factors using the combined asymptotic normalization coefficient (ANC) method [246].

For the $^{14}C(d,p)^{15}C$ reaction, there are three references [74, 75 and 71] with $E_d$=14, 16, and 17 MeV. When data from these references are plotted in Figure 11, they do not agree with each other within a factor of two. However, in each case, the spectroscopic factors quoted in the original references are within 20% of each other (0.88 [74], 0.99 [75], 1.03 [71]). The near agreement of the published SF values, even though the measured cross-sections are very different, illustrates the problem when spectroscopic factors of desired values could be obtained by choosing different input parameters in the analysis. It underscores the importance of analysis with a systematic and consistent approach as studied here.

Since we generally do not analyze data that miss the first forward angle peak, we excluded data taken at 16 (closed squares) and 17 (open circles) MeV [75, 71]. The predicted angular distribution shape (solid curve) shows good agreement with data at 14 MeV [74] at angles less than 15°. We choose to adopt the extracted SF from this data set. However, there may be some problems with the assigned SF value of 1.12. It is about 35% higher compared with the SF's values extracted at low energy and behaves differently compared to other neutron rich nuclei as explained in the next section.

## XI. Dependence of spectroscopic factors on neutron separation energy

Recent measurements of spectroscopic factors from single-nucleon "knock-out" reactions with radioactive and stable nuclei show increasing quenching of the spectroscopic factor values with respect to large-basis shell-model values with nucleon separation energy [247, 248]. The wide range of isotopes studied in this work and listed in Table III includes nuclei with neutron-separation energies ranging from 0.5 to 19 MeV. To examine any quenching trend, we compute the neutron spectroscopic factors using Oxbash, a large-basis shell model code [249, 250]. The model space and the interactions used in the calculations are listed in Table III. Due to the amount of CPU times involved, we cannot compute the SF values from Oxbash for every nucleus. As discussed in detail in ref. [10], excluding the deformed nuclei and nuclei with small SF values, most of the extracted spectroscopic factors agree well with the predicted values from large-basis shell model to 20%.

Figure 12 shows the ratio of the experimental SF values to the LB-SM values from Oxbash as a function of the neutron separation energy. Within the experimental uncertainties, we do not see the systematic quenching of the spectroscopic factors with increasing nucleon separation energy reported for measurements of nucleon knockout reactions induced by radioactive beams. Rather, there seems to be some indication that the trend is the opposite, i.e., the SF values are smaller than the predicted values for nuclei with small neutron separation energy. This trend persists in a smaller subset of the nuclei such as the Ca isotopes plotted as solid stars.

The structures of the neutron rich nuclei with small neutron separation energy are of general interest. For loosely bound nuclei, knockout reactions with radioactive beams suggest no quenching. In our data set, there are seven nuclei with $S_n$<4 MeV, $^{8}$Li, $^{9}$Be, $^{11}$Be, $^{12}$B, $^{15}$C, $^{16}$N, and $^{19}$O. Except for $^{15}$C, which was discussed in previous section, the fits and quality of the data are comparable to that of the other data we have examined. However, the experimental SF values for these nuclei are consistently smaller than the large-basis shell-model predictions. (If we relax the criterion to $S_n$<5 MeV, the conclusion is similar.) Excluding $^{15}$C, the average quenching factor is 0.6. To be sure, we do not have many nuclei and they are all light nuclei with Z≤8. Furthermore, the suppression ratios vary from 0.44 to 0.79 for the six nuclei we examined. In addition, the SF values (as a group) do not agree with the LB-SM predictions. This is somewhat of a puzzle. These results may indicate that the standard global potential [15] may not be appropriate to describe the scattering of these weakly bound nuclei with diffuse surfaces.

Further study with improved theoretical inputs is needed to understand these nuclei with loosely bound neutrons.

**XII. Summary**

In summary, we have evaluated angular distribution measurements from past (p,d) and (d,p) transfer reactions using targets ranging from Li to Cr isotopes. Problems with past measurements are discussed and resolved mainly by comparing to other measurements. Most of these problems concern the consistencies between measurements and would not be limited to (p,d) and (d,p) reactions. Thus the procedure developed to monitor the quality control of the data sets should be applicable to other analysis with large number of data sets. Based on the analysis of the evaluated data and a reaction model with minimum assumptions, we develop a consistent approach to extract spectroscopic factors. Comparisons between spectroscopic factors obtained from (p,d) and (d,p) reactions, suggest that most of the extracted values have uncertainties less than 20%. Thus our SF values have smaller random uncertainties than the values compiled by Endt. Furthermore, the method should be applicable to other stable beams and maybe rare isotope beam experiments. The present compilation of the neutron ground state spectroscopic factors of 80 nuclei provides important reference points for more sophisticated theoretical work on transfer reactions and development in nuclear structure model. For most nuclei, the agreement between data and LB-SM predictions is within 20%. Even though most of the nuclei studied are close to the valley of stability, the nuclei studied here range in neutron separation energy from 0.5 to 19 MeV. The present work does not support the observation that spectroscopic factors are suppressed with increasing neutron separation energy as found in nucleon knockout reactions.

**Acknowledgement**

The authors would like to thank Prof. J. Tostevin for his generosity in giving us the TWOFNR code and helping us use it. We would like to thank Professor K. Kemper for many fruitful discussions and encouragement over the past two years. We acknowledge JINA (Joint Institute of Nuclear Physics) for providing support in creating the web site that contains the digitized and calculated angular distributions for reactions listed in Table I [242]. We acknowledge the support from the Summer for Undergraduate Research Experience (SURE) program at the Chinese University of Hong Kong. This work was supported by the National Science Foundation under Grant No. PHY-01-10253.

Table I: List of reactions studied in this work. SF stands for spectroscopic factors. Not all the SF values extracted are used in computing the averages of the spectroscopic factor for a specific nucleus. The extracted values not used are listed in the 5th column. Most of these include reactions at low beam energy ($E_{beam}$<10 MeV). Those values marked with * are obtained from data which are determined to be problematic. Listed in the last column are abbreviated comments, BS (bad shape), BD (bad data), AU (arbitrary unit), No (Normalization problem), NP (missing first peak), and QV (low Q-values).

| Isotope | Reaction | $E_{beam}$ (MeV) | References | SF(not used) | SF | <SF> | # of points | Comment |
|---|---|---|---|---|---|---|---|---|
| $^{6}$Li | $^{6}$Li(p,d)$^{5}$Li | 33.6 | PR163(1967)1066 | | 1.12 | 1.12 | 3 | |
| $^{7}$Li | $^{6}$Li(d,p)$^{7}$Li | 4.5 | NPA147(1970)65 | 1.59 | | | 2 | |
| $^{7}$Li | $^{6}$Li(d,p)$^{7}$Li | 4.75 | NPA147(1970)65 | 1.81 | | | 2 | |
| $^{7}$Li | $^{6}$Li(d,p)$^{7}$Li | 5 | NPA147(1970)65 | 1.90* | | | 2 | BS |
| $^{7}$Li | $^{6}$Li(d,p)$^{7}$Li | 5.25 | NPA147(1970)65 | 1.78 | | | 3 | |
| $^{7}$Li | $^{6}$Li(d,p)$^{7}$Li | 5.5 | NPA147(1970)65 | 1.70 | | | 3 | |
| $^{7}$Li | $^{6}$Li(d,p)$^{7}$Li | 12 | PR164(1967)1274 | | 1.85 | 1.85 | 2 | |
| $^{7}$Li | $^{7}$Li(p,d)$^{6}$Li | 30.3 | NPA126(1969)261 | 0.34* | | | 3 | BS |
| $^{7}$Li | $^{7}$Li(p,d)$^{6}$Li | 33.6 | PR163(1967)1066 | 0.86* | | | 3 | BS |
| $^{8}$Li | $^{7}$Li(d,p)$^{8}$Li | 12 | PR164(1967)1274 | | 0.62 | 0.62 | 3 | |
| $^{9}$Li | $^{8}$Li(d,p)$^{9}$Li | 19.1 | PRL94(2005)082502 | | 0.98 | 0.98 | 5 | |
| $^{9}$Be | $^{9}$Be(p,d)$^{8}$Be | 5 | NPA184(1972)175 | 0.43 | | | 7 | |
| $^{9}$Be | $^{9}$Be(p,d)$^{8}$Be | 6 | NPA184(1972)175 | 0.47 | | | 4 | |
| $^{9}$Be | $^{9}$Be(p,d)$^{8}$Be | 7 | NPA184(1972)175 | 0.45 | | | 3 | |
| $^{9}$Be | $^{9}$Be(p,d)$^{8}$Be | 8 | NPA184(1972)175 | 0.51 | | | 3 | |
| $^{9}$Be | $^{9}$Be(p,d)$^{8}$Be | 9 | NPA184(1972)175 | 0.53 | | | 2 | |
| $^{9}$Be | $^{9}$Be(p,d)$^{8}$Be | 10 | NPA184(1972)175 | | 0.46 | | 2 | BS |
| $^{9}$Be | $^{9}$Be(p,d)$^{8}$Be | 11 | NPA184(1972)175 | | 0.46 | | 2 | BS |
| $^{9}$Be | $^{9}$Be(p,d)$^{8}$Be | 14.3 | PRC24(1981)2401 | | 0.41 | | 2 | BS |
| $^{9}$Be | $^{9}$Be(p,d)$^{8}$Be | 15 | NPA157(1970)305 | | 0.42 | | 3 | BS |
| $^{9}$Be | $^{9}$Be(p,d)$^{8}$Be | 17 | NPA199(1973)433 | | 0.51 | | 3 | BS |
| $^{9}$Be | $^{9}$Be(p,d)$^{8}$Be | 21 | NPA199(1973)433 | | 0.50 | | 2 | BS |
| $^{9}$Be | $^{9}$Be(p,d)$^{8}$Be | 25 | NPA199(1973)433 | | 0.43 | | 2 | BS |
| $^{9}$Be | $^{9}$Be(p,d)$^{8}$Be | 26.2 | PRC24(1981)2401 | 0.35* | | | 1 | BS |
| $^{9}$Be | $^{9}$Be(p,d)$^{8}$Be | 29.1 | NPA199(1973)433 | | 0.48 | | 2 | BS |
| $^{9}$Be | $^{9}$Be(p,d)$^{8}$Be | 33.6 | PR163(1967)1066 | | 0.44 | | 1 | BS |
| $^{9}$Be | $^{9}$Be(p,d)$^{8}$Be | 46 | PR153(1967)1127 | | 0.49 | 0.46 | 1 | BS |
| $^{10}$Be | $^{9}$Be(d,p)$^{10}$Be | 4.5 | NPA147(1970)65 | 2.44 | | | 2 | |
| $^{10}$Be | $^{9}$Be(d,p)$^{10}$Be | 4.75 | NPA147(1970)65 | 2.11 | | | 3 | |
| $^{10}$Be | $^{9}$Be(d,p)$^{10}$Be | 5 | NPA147(1970)65 | 2.14 | | | 2 | |
| $^{10}$Be | $^{9}$Be(d,p)$^{10}$Be | 5.25 | NPA147(1970)65 | 2.06 | | | 3 | |
| $^{10}$Be | $^{9}$Be(d,p)$^{10}$Be | 5.5 | NPA147(1970)65 | 2.01 | | | 2 | |
| $^{10}$Be | $^{9}$Be(d,p)$^{10}$Be | 5.75 | NPA147(1970)65 | 1.83 | | | 3 | |
| $^{10}Be$ | $^{9}Be(d,p)^{10}Be$ | 6 | NPA147(1970)65 | 2.01 | | | 3 | |
| $^{10}Be$ | $^{9}Be(d,p)^{10}Be$ | 6.5 | J,IZV,64,440,2000 | 1.55 | | | 5 | |
| $^{10}Be$ | $^{9}Be(d,p)^{10}Be$ | 7 | J,IZV,64,440,2000 | 1.48 | | | 4 | |
| $^{10}Be$ | $^{9}Be(d,p)^{10}Be$ | 7.5 | J,IZV,64,440,2000 | 1.07 | | | 2 | |
| $^{10}Be$ | $^{9}Be(d,p)^{10}Be$ | 8 | J,IZV,64,440,2000 | 1.05 | | | 1 | |
| $^{10}Be$ | $^{9}Be(d,p)^{10}Be$ | 8.5 | J,IZV,64,440,2000 | 1.11 | | | 2 | |
| $^{10}Be$ | $^{9}Be(d,p)^{10}Be$ | 9 | J,IZV,64,440,2000 | 1.10 | | | 2 | |
| $^{10}Be$ | $^{9}Be(d,p)^{10}Be$ | 9.5 | J,IZV,64,440,2000 | 1.03 | | | 2 | |
| $^{10}Be$ | $^{9}Be(d,p)^{10}Be$ | 10 | J,IZV,64,440,2000 | 1.10* | | | 2 | NP |
| $^{10}Be$ | $^{9}Be(d,p)^{10}Be$ | 10.5 | J,IZV,64,440,2000 | 1.18* | | | 2 | NP |
| $^{10}Be$ | $^{9}Be(d,p)^{10}Be$ | 11 | J,IZV,64,440,2000 | 1.17* | | | 2 | BD |
| $^{10}Be$ | $^{9}Be(d,p)^{10}Be$ | 11.8 | PR164(1967)1270 | | 1.49 | | 3 | BD |
| $^{10}Be$ | $^{9}Be(d,p)^{10}Be$ | 11.8 | NPA53(1964)77 | | 1.42 | | 2 | |
| $^{10}Be$ | $^{9}Be(d,p)^{10}Be$ | 12.5 | J,YF,45,312,(1987) | | 1.72 | | 4 | NP |
| $^{10}Be$ | $^{9}Be(d,p)^{10}Be$ | 15 | NPA266(1976)29 | | 1.75 | | 4 | |
| $^{10}Be$ | $^{9}Be(d,p)^{10}Be$ | 15.3 | J,YF,64,1995,2001 | | 1.40 | 1.58 | 4 | NP |
| $^{10}Be$ | $^{9}Be(d,p)^{10}Be$ | 17.3 | NPA236(1974)77 | 0.99* | | | 3 | BS |
| $^{10}Be$ | $^{9}Be(d,p)^{10}Be$ | 28 | PR126(1962)1059 | 2.26* | | | 2 | BS |
| $^{10}Be$ | $^{10}Be(p,d)^{9}Be$ | 49.8 | Liu, thesis (2005) | 2.96* | | | 10 | BD |
| $^{11}Be$ | $^{10}Be(d,p)^{11}Be$ | 12 | NPA157(1970)305 | | 0.44 | | 3 | |
| $^{11}Be$ | $^{10}Be(d,p)^{11}Be$ | 25 | NPA315(1979)124 | | 0.53 | 0.49 | 3 | |
| $^{11}Be$ | $^{11}Be(p,d)^{10}Be$ | 35.3 | NPA683(2001)48 | | 0.57 | 0.57 | 2 | |
| $^{10}B$ | $^{10}B(p,d)^{9}B$ | 33.6 | PR167(1968)963 | | 0.57 | | 4 | |
| $^{10}B$ | $^{10}B(p,d)^{9}B$ | 49.5 | NPA141(1970)158 | | 0.43 | 0.50 | 3 | |
| $^{11}B$ | $^{10}B(d,p)^{11}B$ | 4.5 | NPA147(1970)65 | 1.11 | | | 2 | |
| $^{11}B$ | $^{10}B(d,p)^{11}B$ | 4.75 | NPA147(1970)65 | 1.06 | | | 3 | |
| $^{11}B$ | $^{10}B(d,p)^{11}B$ | 5 | NPA147(1970)65 | 0.92 | | | 2 | |
| $^{11}B$ | $^{10}B(d,p)^{11}B$ | 5.25 | NPA147(1970)65 | 0.85 | | | 2 | |
| $^{11}B$ | $^{10}B(d,p)^{11}B$ | 5.5 | NPA147(1970)65 | 0.81 | | | 2 | |
| $^{11}B$ | $^{10}B(d,p)^{11}B$ | 8.2 | PR131(1963)304 | 5.05 | | | 3 | AU |
| $^{11}B$ | $^{10}B(d,p)^{11}B$ | 10.1 | NP38(1962)114 | 1.00* | | | 4 | BD |
| $^{11}B$ | $^{10}B(d,p)^{11}B$ | 12 | PR164(1967)1274 | | 1.25 | | 2 | BS |
| $^{11}B$ | $^{10}B(d,p)^{11}B$ | 13.5 | NP73(1965)473 | | 1.68 | | 5 | |
| $^{11}B$ | $^{10}B(d,p)^{11}B$ | 15.5 | PR131(1963)304 | 1.50* | | | 6 | AU |
| $^{11}B$ | $^{10}B(d,p)^{11}B$ | 21.5 | PR131(1963)304 | 0.32* | | | 9 | AU |
| $^{11}B$ | $^{10}B(d,p)^{11}B$ | 28 | PR126(1962)1059 | | 1.52 | 1.55 | 2 | |
| $^{11}B$ | $^{10}B(d,p)^{11}B$ | 28 | PR131(1963)304 | 0.06* | | | | AU |
| $^{11}B$ | $^{11}B(p,d)^{10}B$ | 19 | PR129(1963)272 | 3.16* | | | 3 | BD |
| $^{11}B$ | $^{11}B(p,d)^{10}B$ | 33.6 | PR167(1968)963 | | 1.29 | 1.29 | 3 | |
| $^{11}B$ | $^{11}B(p,d)^{10}B$ | 44.1 | Liu, thesis (2005) | 1.05* | | | 2 | BD |
| $^{12}B$ | $^{11}B(d,p)^{12}B$ | 11.8 | PRC64(2001)034312 | | 0.44 | | 5 | |
| $^{12}B$ | $^{11}B(d,p)^{12}B$ | 12 | J,NUK,19,693,1974 | | 0.47 | 0.45 | 3 | |
| $^{12}B$ | $^{11}B(d,p)^{12}B$ | 12 | PR164(1967)1274 | 0.35* | | | 1 | BS |
| $^{12}C$ | $^{12}C(p,d)^{11}C$ | 19.3 | PR128(1962)1810 | | | | | QV |
| $^{12}C$ | $^{12}C(p,d)^{11}C$ | 19.5 | PR128(1962)1810 | | | | | QV |
| $^{12}C$ | $^{12}C(p,d)^{11}C$ | 20 | PR128(1962)1810 | | | | | QV |
| $^{12}C$ | $^{12}C(p,d)^{11}C$ | 30.3 | NPA99(1967)669 | | 2.68 | | 3 | |
| $^{12}C$ | $^{12}C(p,d)^{11}C$ | 39.8 | PR132(1963)813 | 5.50* | | | 4 | No |
| $^{12}C$ | $^{12}C(p,d)^{11}C$ | 61 | PRC21(1980)2162 | | 3.36 | | 6 | |
| $^{12}C$ | $^{12}C(p,d)^{11}C$ | 65 | NPA255(1975)187 | | 3.07 | 3.12 | 3 | |
| $^{12}C$ | $^{12}C(p,d)^{11}C$ | 65 | NPA343(1980)234 | 3.03* | | | 1 | BS |
| $^{13}C$ | $^{12}C(d,p)^{13}C$ | 4 | NP82(1966)161 | 0.64 | | | 3 | |
| $^{13}C$ | $^{12}C(d,p)^{13}C$ | 4.5 | NP82(1966)161 | 0.67 | | | 2 | |
| $^{13}C$ | $^{12}C(d,p)^{13}C$ | 4.5 | PR101(1956)209 | 0.59 | | | 2 | |
| $^{13}C$ | $^{12}C(d,p)^{13}C$ | 4.5 | NPA129(1969)405 | 0.43 | | | 2 | |
| $^{13}C$ | $^{12}C(d,p)^{13}C$ | 7.15 | JETP12(1960)1 | 0.88 | | | 4 | |
| $^{13}C$ | $^{12}C(d,p)^{13}C$ | 8.9 | NP22(1961)34 | 0.92 | | | 6 | |
| $^{13}C$ | $^{12}C(d,p)^{13}C$ | 10.2 | PR123(1961)619 | | 0.85 | | 3 | |
| $^{13}C$ | $^{12}C(d,p)^{13}C$ | 11.8 | NP53(1964)77 | | 0.82 | | 3 | |
| $^{13}C$ | $^{12}C(d,p)^{13}C$ | 11.8 | PRC64(2001)034312 | 0.60* | | | 2 | BD |
| $^{13}C$ | $^{12}C(d,p)^{13}C$ | 12 | NPA477(1988)77 | | 0.71 | | 2 | |
| $^{13}C$ | $^{12}C(d,p)^{13}C$ | 12 | PR164(1967)1274 | | 0.87 | | 3 | |
| $^{13}C$ | $^{12}C(d,p)^{13}C$ | 12.4 | PR123(1961)619 | | 0.78 | | 4 | |
| $^{13}C$ | $^{12}C(d,p)^{13}C$ | 14.7 | PR123(1961)619 | | 0.72 | | 3 | |
| $^{13}C$ | $^{12}C(d,p)^{13}C$ | 14.8 | PR100(1955)235 | | 0.77 | | 1 | |
| $^{13}C$ | $^{12}C(d,p)^{13}C$ | 15 | NPA208(1973)77 | | 0.68 | | 2 | |
| $^{13}C$ | $^{12}C(d,p)^{13}C$ | 16.6 | JPSJ 15(1960)550 | | 0.59 | | 2 | |
| $^{13}C$ | $^{12}C(d,p)^{13}C$ | 19.6 | JPSJ 15(1960)550 | | 0.61 | | 2 | |
| $^{13}C$ | $^{12}C(d,p)^{13}C$ | 25.9 | PR136(1964)B1682 | | 0.66 | | 6 | |
| $^{13}C$ | $^{12}C(d,p)^{13}C$ | 30 | NPA448(86)205 | | 0.62 | 0.73 | 2 | BS |
| $^{13}C$ | $^{12}C(d,p)^{13}C$ | 51 | Liu, thesis (2005) | | | | | BD |
| $^{13}C$ | $^{12}C(d,p)^{13}C$ | 56 | NPA419(84)530 | 0.99* | | | 1 | NP |
| $^{13}C$ | $^{13}C(p,d)^{12}C$ | 35 | PRC51(1995)2592 | | 0.79 | | 2 | BS |
| $^{13}C$ | $^{13}C(p,d)^{12}C$ | 41.3 | NPA470(1987)349 | | 0.86 | | 1 | BS |
| $^{13}C$ | $^{13}C(p,d)^{12}C$ | 48.3 | Liu, thesis (2005) | | 0.90 | | 5 | BS |
| $^{13}C$ | $^{13}C(p,d)^{12}C$ | 55 | PLB27(1968)625 | | 0.67 | 0.81 | 3 | BS |
| $^{13}C$ | $^{13}C(p,d)^{12}C$ | 65 | NPA343(1980)234 | 1.61* | | | 3 | NP |
| $^{14}C$ | $^{13}C(d,p)^{14}C$ | 12 | PR164(1967)1274 | | 1.94 | | 5 | |
| $^{14}C$ | $^{13}C(d,p)^{14}C$ | 13 | NPA312(1978)1 | | 1.61 | 1.82 | 3 | NP |
| $^{14}C$ | $^{13}C(d,p)^{14}C$ | 56 | NPA419(1984)530 | 2.34* | | | 2 | NP |
| $^{14}C$ | $^{14}C(p,d)^{13}C$ | 14.5 | NPA165(1971)19 | 0.88* | | | 4 | NP |
| $^{14}C$ | $^{14}C(p,d)^{13}C$ | 18.5 | PR129(1963)272 | | 1.87 | | 3 | |
| $^{14}C$ | $^{14}C(p,d)^{13}C$ | 27 | NPA255(1975)243 | | 1.02 | | 4 | |
| $^{14}C$ | $^{14}C(p,d)^{13}C$ | 35 | NPA509(1990)141 | | 1.66 | 1.50 | 5 | |
| $^{15}C$ | $^{14}C(d,p)^{15}C$ | 2 | NPA96(1967) 671 | 1.07 | | | 2 | |
| $^{15}C$ | $^{14}C(d,p)^{15}C$ | 2.6 | NPA96(1967) 671 | 0.66 | | | 1 | |
| $^{15}C$ | $^{14}C(d,p)^{15}C$ | 3 | NPA96(1967) 671 | 0.73 | | | 2 | |
| $^{15}C$ | $^{14}C(d,p)^{15}C$ | 3.4 | NPA96(1967) 671 | 0.78 | | | 2 | |
| $^{15}C$ | $^{14}C(d,p)^{15}C$ | 14 | PRC12(1975)1730 | | 1.12 | 1.12 | 1 | |
| $^{15}C$ | $^{14}C(d,p)^{15}C$ | 16 | NPA579(1994)125 | 1.15* | | | | NP |
| $^{15}C$ | $^{14}C(d,p)^{15}C$ | 17 | NPA255(1975)243 | 0.42* | | | | BS |
| $^{14}N$ | $^{14}N(p,d)^{13}N$ | 14.5 | NPA165(1971)19 | | 0.68 | | 5 | |
| $^{14}N$ | $^{14}N(p,d)^{13}N$ | 18.5 | PR122(1961)595 | | 0.76 | | 3 | |
| $^{14}N$ | $^{14}N(p,d)^{13}N$ | 21 | NPA322(1979)117 | 0.60* | | | 2 | NP |
| $^{14}N$ | $^{14}N(p,d)^{13}N$ | 30.3 | NPA99(1967)540 | | 1.00 | 0.77 | 2 | |
| $^{14}N$ | $^{14}N(p,d)^{13}N$ | 65 | NPA255(1975)187 | 0.48* | | | 2 | NP |
| $^{15}N$ | $^{14}N(d,p)^{15}N$ | 10 | NPA333(1980)13 | | | | | BD |
| $^{15}N$ | $^{14}N(d,p)^{15}N$ | 10.03 | NPA382(1982)269 | | 1.66 | | 2 | |
| $^{15}N$ | $^{14}N(d,p)^{15}N$ | 11.65 | NPA382(1982)269 | | | | | NP |
| $^{15}N$ | $^{14}N(d,p)^{15}N$ | 12 | PR164(1967)1274 | | 1.12 | | 3 | |
| $^{15}N$ | $^{14}N(d,p)^{15}N$ | 14.8 | PR105(1957)639 | | 1.58 | | 5 | |
| $^{15}N$ | $^{14}N(d,p)^{15}N$ | 31 | NPA275(1977)212 | | 1.18 | 1.39 | 3 | |
| $^{15}N$ | $^{14}N(d,p)^{15}N$ | 52 | NPA275(1977)212 | 1.94* | | | | BD |
| $^{15}N$ | $^{15}N(p,d)^{14}N$ | 18.6 | PR122(1961)595 | | 1.76 | | 4 | |
| $^{15}N$ | $^{15}N(p,d)^{14}N$ | 39.8 | PR187(1969)1259 | | 1.43 | 1.65 | 2 | |
| $^{16}N$ | $^{15}N(d,p)^{16}N$ | 14.8 | PR105(1957)639 | | 0.42 | 0.42 | 4 | |
| $^{16}O$ | $^{16}O(p,d)^{15}O$ | 18.5 | PR129(1963)272 | 1.74* | | | 4 | BS |
| $^{16}O$ | $^{16}O(p,d)^{15}O$ | 19 | PR129(1963)272 | 2.33* | | | 5 | BS |
| $^{16}O$ | $^{16}O(p,d)^{15}O$ | 20 | PR129(1963)272 | | 2.32 | | 4 | |
| $^{16}O$ | $^{16}O(p,d)^{15}O$ | 21.27 | PR187(1969)1246 | 1.69* | | | 5 | |
| $^{16}O$ | $^{16}O(p,d)^{15}O$ | 25.52 | PR187(1969)1246 | | 2.82 | | 4 | |
| $^{16}O$ | $^{16}O(p,d)^{15}O$ | 30.3 | NPA99(1967)669 | | 2.31 | | 4 | |
| $^{16}O$ | $^{16}O(p,d)^{15}O$ | 31.82 | PR187(1969)1246 | | 2.29 | | 2 | |
| $^{16}O$ | $^{16}O(p,d)^{15}O$ | 38.63 | PR187(1969)1246 | | 2.09 | | 4 | |
| $^{16}O$ | $^{16}O(p,d)^{15}O$ | 39.8 | PR132(1963)813 | | 2.59 | | 2 | |
| $^{16}O$ | $^{16}O(p,d)^{15}O$ | 45.34 | PR187(1969)1246 | | 2.70 | | 4 | |
| $^{16}O$ | $^{16}O(p,d)^{15}O$ | 65 | NPA255(1975)187 | 2.32* | | 2.46 | 1 | NP |
| $^{16}O$ | $^{16}O(p,d)^{15}O$ | 65 | NPA343(1980)234 | 2.75* | | | 1 | NP |
| $^{17}O$ | $^{16}O(d,p)^{17}O$ | 1.3 | NP82(1966)161 | | | | | |
| $^{17}O$ | $^{16}O(d,p)^{17}O$ | 2.279 | NPA114(1968)330 | | | | | |
| $^{17}O$ | $^{16}O(d,p)^{17}O$ | 2.582 | NPA114(1968)330 | 1.54 | | | 1 | |
| $^{17}O$ | $^{16}O(d,p)^{17}O$ | 2.864 | NPA114(1968)330 | 1.54 | | | 1 | |
| $^{17}O$ | $^{16}O(d,p)^{17}O$ | 3.155 | NPA114(1968)330 | 1.56 | | | 1 | |
| $^{17}O$ | $^{16}O(d,p)^{17}O$ | 3.49 | PR123(1961)619 | 2.57 | | | 2 | |
| $^{17}O$ | $^{16}O(d,p)^{17}O$ | 4 | NP82(1966)161 | 2.39 | | | 4 | |
| $^{17}O$ | $^{16}O(d,p)^{17}O$ | 4.11 | PR123(1961)619 | 2.11 | | | 2 | |
| $^{17}O$ | $^{16}O(d,p)^{17}O$ | 6 | NPA112(1968)76 | 1.24 | | | 6 | |
| $^{17}O$ | $^{16}O(d,p)^{17}O$ | 6.26 | NPA134(1969)561 | 1.17 | | | 3 | |
| $^{17}O$ | $^{16}O(d,p)^{17}O$ | 7.5 | NPA112(1968)76 | 1.26 | | | 6 | |
| $^{17}O$ | $^{16}O(d,p)^{17}O$ | 7.85 | NPA112(1968)76 | 1.22 | | | 6 | |
| $^{17}$O | $^{16}$O(d,p)$^{17}$O | 8 | NPA172(1971)663 | 1.40 | | | 1 | |
| $^{17}$O | $^{16}$O(d,p)$^{17}$O | 8.2 | NPA112(1968)76 | 1.11 | | | 6 | |
| $^{17}$O | $^{16}$O(d,p)$^{17}$O | 8.55 | NPA112(1968)76 | 0.96 | | | 6 | |
| $^{17}$O | $^{16}$O(d,p)$^{17}$O | 9 | PR123(1961)619 | 0.98 | | | 3 | |
| $^{17}$O | $^{16}$O(d,p)$^{17}$O | 9.3 | NPA188(1972)164 | 0.88 | | | 3 | |
| $^{17}$O | $^{16}$O(d,p)$^{17}$O | 10 | NPA112(1968)76 | | 1.04 | | 3 | |
| $^{17}$O | $^{16}$O(d,p)$^{17}$O | 10.2 | PR123(1961)619 | | 0.78 | | 2 | BD |
| $^{17}$O | $^{16}$O(d,p)$^{17}$O | 11 | NPA112(1968)76 | | 0.88 | | 2 | |
| $^{17}$O | $^{16}$O(d,p)$^{17}$O | 11.8 | NP53(1964)77 | 0.62* | | | 3 | BS |
| $^{17}$O | $^{16}$O(d,p)$^{17}$O | 12 | NPA97(1967)541 | 0.47* | | | 4 | BD |
| $^{17}$O | $^{16}$O(d,p)$^{17}$O | 12.4 | PR123(1961)619 | | 1.03 | | 3 | |
| $^{17}$O | $^{16}$O(d,p)$^{17}$O | 13.3 | NPA188(1972)164 | | 1.13 | | 5 | |
| $^{17}$O | $^{16}$O(d,p)$^{17}$O | 14.8 | PR123(1961)619 | | 0.98 | | 2 | |
| $^{17}$O | $^{16}$O(d,p)$^{17}$O | 15 | PR121(1961)820 | | 1.02 | | 3 | |
| $^{17}$O | $^{16}$O(d,p)$^{17}$O | 19 | PR123(1961)619 | 0.81* | | | 1 | BS |
| $^{17}$O | $^{16}$O(d,p)$^{17}$O | 25.4 | NPA218(1974)249 | | 0.89 | | 3 | |
| $^{17}$O | $^{16}$O(d,p)$^{17}$O | 26.3 | NP50(1964)479 | 1.37* | | | 6 | |
| $^{17}$O | $^{16}$O(d,p)$^{17}$O | 36 | NPA218(1974)249 | | 0.87 | | 4 | |
| $^{17}$O | $^{16}$O(d,p)$^{17}$O | 63.2 | NPA218(1974)249 | | 1.07 | 0.99 | 3 | |
| $^{17}$O | $^{17}$O(p,d)$^{16}$O | 8.62 | NPA244(1975)77 | 1.10 | | | 4 | |
| $^{17}$O | $^{17}$O(p,d)$^{16}$O | 9.56 | NPA244(1975)77 | 1.01 | | | 2 | BS |
| $^{17}$O | $^{17}$O(p,d)$^{16}$O | 10.5 | NPA244(1975)77 | | 0.78 | | 4 | |
| $^{17}$O | $^{17}$O(p,d)$^{16}$O | 11.16 | NPA244(1975)77 | 0.70* | | | 2 | BS |
| $^{17}$O | $^{17}$O(p,d)$^{16}$O | 11.44 | NPA244(1975)77 | | 0.74 | | 4 | |
| $^{17}$O | $^{17}$O(p,d)$^{16}$O | 31 | PLB31(1970)126 | | 0.99 | 0.81 | 2 | |
| $^{18}$O | $^{17}$O(d,p)$^{18}$O | 18 | PRC13(1976)55 | | 1.80 | 1.80 | 3 | |
| $^{18}$O | $^{18}$O(p,d)$^{17}$O | 17.6 | PR129(1963)272 | | 1.72 | | 4 | |
| $^{18}$O | $^{18}$O(p,d)$^{17}$O | 18.2 | NPA101(1967)241 | | 1.43 | 1.60 | 3 | |
| $^{18}$O | $^{18}$O(p,d)$^{17}$O | 20 | PRC10(1974)445 | 0.79* | | | 2 | BS |
| $^{18}$O | $^{18}$O(p,d)$^{17}$O | 24.4 | PRC10(1974)445 | 1.50* | | | 2 | BS |
| $^{18}$O | $^{18}$O(p,d)$^{17}$O | 29.8 | PRC10(1974)445 | 1.40* | | | 3 | BS |
| $^{18}$O | $^{18}$O(p,d)$^{17}$O | 37.5 | PRC10(1974)445 | 0.97* | | | 1 | NP |
| $^{18}$O | $^{18}$O(p,d)$^{17}$O | 43.6 | PRC10(1974)445 | 1.01* | | | 2 | BD |
| $^{19}$O | $^{18}$O(d,p)$^{19}$O | 10 | NPA331(1979)269 | 0.63* | | | 1 | NP |
| $^{19}$O | $^{18}$O(d,p)$^{19}$O | 14.8 | NPA219(1974)429 | | 0.47 | | 4 | |
| $^{19}$O | $^{18}$O(d,p)$^{19}$O | 15 | PR122(1961)150 | | 0.38 | 0.43 | 3 | |
| $^{19}$F | $^{19}$F(p,d)$^{18}$F | 18.5 | PR122(1961)595 | | 1.62 | | 4 | |
| $^{19}$F | $^{19}$F(p,d)$^{18}$F | 19.3 | NPA337(1980)107 | | 1.58 | 1.60 | 3 | |
| $^{20}$F | $^{19}$F(d,p)$^{20}$F | 12 | PRC10(1974)1292 | | 0.013 | 0.013 | 3 | |
| $^{20}$F | $^{19}$F(d,p)$^{20}$F | 16 | PRC6(1972)21 | | | | | BD |
| $^{21}$Ne | $^{20}$Ne(d,p)$^{21}$Ne | 11 | NPA332(1979)125 | | 0.044 | | 2 | |
| $^{21}$Ne | $^{20}$Ne(d,p)$^{21}$Ne | 16.4 | NPA152(1970)317 | | 0.031 | 0.035 | 5 | |
| $^{21}$Ne | $^{21}$Ne(p,d)$^{20}$Ne | 20 | NPA150(1970)609 | | 0.03 | 0.03 | 8 | BS |
| $^{22}$Ne | $^{21}$Ne(d,p)$^{22}$Ne | 10.2 | NPA150(1970)609 | | | | | BD |
| $^{22}$Ne | $^{22}$Ne(p,d)$^{21}$Ne | 18.2 | NPA95(1967)591 | | 0.26 | | 4 | |
| $^{22}$Ne | $^{22}$Ne(p,d)$^{21}$Ne | 20 | PR184(1969)1094 | | 0.20 | 0.24 | 2 | |
| $^{23}$Ne | $^{22}$Ne(d,p)$^{23}$Ne | 12.1 | NPA152(1970)317 | | 0.24 | | 6 | |
| $^{23}$Ne | $^{22}$Ne(d,p)$^{23}$Ne | 12.1 | NPA95(1967)591 | | 0.24 | 0.24 | 6 | |
| $^{24}$Na | $^{23}$Na(d,p)$^{24}$Na | 7.83 | NP45(1963)273 | | 0.59 | 0.59 | 2 | |
| $^{24}$Mg | $^{24}$Mg(p,d)$^{23}$Mg | 27.3 | PR177(1969)1737 | | 0.39 | | 4 | |
| $^{24}$Mg | $^{24}$Mg(p,d)$^{23}$Mg | 33.6 | PR172(1968)1078 | 0.34* | | | 2 | BD |
| $^{24}$Mg | $^{24}$Mg(p,d)$^{23}$Mg | 49.2 | PRC33(1986)22 | | 0.44 | 0.41 | 3 | |
| $^{25}$Mg | $^{24}$Mg(d,p)$^{25}$Mg | 5 | NP88(1966)654 | 0.75 | | | 6 | |
| $^{25}$Mg | $^{24}$Mg(d,p)$^{25}$Mg | 6 | NP88(1966)654 | 0.50 | | | 3 | |
| $^{25}$Mg | $^{24}$Mg(d,p)$^{25}$Mg | 10 | NPA203(1973)177 | | 0.28 | | 3 | |
| $^{25}$Mg | $^{24}$Mg(d,p)$^{25}$Mg | 12 | NPA249(1975)205 | | 0.33 | | 3 | BS |
| $^{25}$Mg | $^{24}$Mg(d,p)$^{25}$Mg | 14 | PRC10(1974)556 | | 0.27 | | 3 | |
| $^{25}$Mg | $^{24}$Mg(d,p)$^{25}$Mg | 15 | PRC10(1974)556 | 0.28* | | 0.29 | 1 | BS |
| $^{25}$Mg | $^{24}$Mg(d,p)$^{25}$Mg | 56 | NPA419(530)530 | 0.49* | | | 6 | NP |
| $^{26}$Mg | $^{25}$Mg(d,p)$^{26}$Mg | 8 | NP88(1966)513 | 2.97 | | | 7 | |
| $^{26}$Mg | $^{25}$Mg(d,p)$^{26}$Mg | 12 | PRC29(1984)2013 | | 2.01 | 2.01 | 8 | |
| $^{26}$Mg | $^{25}$Mg(d,p)$^{26}$Mg | 13 | NPA430(1984)234 | 2.62* | | | 7 | BD |
| $^{26}$Mg | $^{26}$Mg(p,d)$^{25}$Mg | 20 | NPA172(1971)99 | | 2.01 | | 2 | |
| $^{26}$Mg | $^{26}$Mg(p,d)$^{25}$Mg | 23.95 | NPA351(1981)77 | | 3.06 | | 4 | |
| $^{26}$Mg | $^{26}$Mg(p,d)$^{25}$Mg | 35 | PRC38(1988)2026 | | 2.97 | 2.80 | 3 | BS |
| $^{27}$Mg | $^{26}$Mg(d,p)$^{27}$Mg | 5.07 | PR136(1964)B1703 | 1.03 | | | 1 | |
| $^{27}$Mg | $^{26}$Mg(d,p)$^{27}$Mg | 12 | NPA230(1974)317 | | 0.45 | 0.45 | 2 | |
| $^{27}$Al | $^{27}$Al(p,d)$^{26}$Al | 20 | NPA204(1973)609 | | 1.51 | | 3 | |
| $^{27}$Al | $^{27}$Al(p,d)$^{26}$Al | 35 | NPA263(1976)293 | | 1.32 | 1.40 | 4 | |
| $^{28}$Al | $^{27}$Al(d,p)$^{28}$Al | 6 | NPA197(1972)97 | 0.43 | | | 3 | |
| $^{28}$Al | $^{27}$Al(d,p)$^{28}$Al | 12 | NPA173(1971)414 | | 0.60 | | 3 | |
| $^{28}$Al | $^{27}$Al(d,p)$^{28}$Al | 23 | PRC5(1972)1313 | | 0.82 | 0.66 | 1 | |
| $^{28}$Si | $^{28}$Si(p,d)$^{27}$Si | 27.6 | NPA107(1968)659 | 15.44* | | | 6 | |
| $^{28}$Si | $^{28}$Si(p,d)$^{27}$Si | 33.6 | PR172(1968)1078 | | 4.40 | 4.40 | 4 | |
| $^{29}$Si | $^{28}$Si(d,p)$^{29}$Si | 5 | NPA120(1968)94 | 0.73 | | | 1 | |
| $^{29}$Si | $^{28}$Si(d,p)$^{29}$Si | 5.8 | NPA149(1970)605 | 0.41 | | | 2 | |
| $^{29}$Si | $^{28}$Si(d,p)$^{29}$Si | 9 | NPA172(1971)663 | 0.39 | | | 1 | |
| $^{29}$Si | $^{28}$Si(d,p)$^{29}$Si | 10 | NPA189(1972)305 | | 0.56 | | 2 | |
| $^{29}$Si | $^{28}$Si(d,p)$^{29}$Si | 17.85 | NPA408(1983)221 | | 0.36 | | 2 | |
| $^{29}$Si | $^{28}$Si(d,p)$^{29}$Si | 18 | PRC4(1971)1778 | | 0.24 | 0.42 | 1 | |
| $^{29}$Si | $^{29}$Si(p,d)$^{28}$Si | 27.3 | PRC2(1970)1440 | 1.32* | | | 2 | NP |
| $^{30}$Si | $^{29}$Si(d,p)$^{30}$Si | 10 | NPA211(1973)7 | | 0.93 | | 1 | BS |
| $^{30}$Si | $^{29}$Si(d,p)$^{30}$Si | 12.3 | NPA468(1987)357 | | | | | NP |
| $^{30}$Si | $^{29}$Si(d,p)$^{30}$Si | 16 | NPA202(1973)497 | | 0.64 | 0.79 | 1 | |
| $^{30}$Si | $^{30}$Si(p,d)$^{29}$Si | 27 | NPA241(1975)285 | | 0.87 | | 3 | |
| $^{30}$Si | $^{30}$Si(p,d)$^{29}$Si | 27.3 | PRC2(1970)1440 | 0.87* | | 0.87 | 1 | NP |
| $^{31}$Si | $^{30}$Si(d,p)$^{31}$Si | 7 | NPA108(1968)49 | 0.58 | | | 5 | |
| $^{31}$Si | $^{30}$Si(d,p)$^{31}$Si | 10 | NPA108(1968)49 | | 0.55 | | 4 | |
| $^{31}$Si | $^{30}$Si(d,p)$^{31}$Si | 12.3 | NPA468(1987)357 | | 0.71 | | 2 | |
| $^{31}$Si | $^{30}$Si(d,p)$^{31}$Si | 12.3 | NPA662(2000)112 | | 0.47 | 0.54 | 6 | |
| $^{32}$P | $^{31}$P(d,p$^{)32}$P | 10 | NPA210(1973)29 | | 0.68 | | 2 | |
| $^{32}$P | $^{31}$P(d,p$^{)32}$P | 20 | NPA501(1989)413 | | 0.48 | 0.58 | 2 | |
| $^{32}$S | $^{32}$S(p,d)$^{31}$S | 24.5 | NPA177(1971)205 | 3.4* | | | 1 | NP |
| $^{32}$S | $^{32}$S(p,d)$^{31}$S | 33.6 | PR172(1968)1078 | | 1.51 | 1.51 | 2 | NP |
| $^{33}$S | $^{32}$S(d,p)$^{33}$S | 18 | PRC4(1971)1778 | | 0.70 | 0.70 | 4 | |
| $^{34}$S | $^{33}$S(d,p)$^{34}$S | 12 | NPA173(1971)456 | | 1.85 | | 4 | |
| $^{34}$S | $^{33}$S(d,p)$^{34}$S | 12 | NPA198(1972)209 | | 1.23 | 1.58 | 3 | |
| $^{34}$S | $^{34}$S(p,d)$^{33}$S | 24.5 | NPA177(1971)205 | | 1.08 | 1.08 | 3 | |
| $^{34}$S | $^{34}$S(p,d)$^{33}$S | 35 | PRC11(1975)654 | 3.30* | | | 8 | BS |
| $^{35}$S | $^{34}$S(d,p)$^{35}$S | 10 | NPA170(1971)607 | | 0.30 | 0.30 | 5 | |
| $^{35}$S | $^{34}$S(d,p)$^{35}$S | 11.8 | NPA287(1977)94 | | 0.30 | | 2 | BS |
| $^{37}$S | $^{36}$S(d,p)$^{37}$S | 12.3 | NPA414(1984)219 | | 0.88 | | 4 | |
| $^{37}$S | $^{36}$S(d,p)$^{37}$S | 25 | PRC30(1984)1442 | | 0.89 | 0.88 | 1 | |
| $^{35}$Cl | $^{35}$Cl(p,d)$^{34}$Cl | 40 | NPA189(1972)513 | | 0.35 | 0.35 | 4 | |
| $^{36}$Cl | $^{35}$Cl(d,p)$^{36}$Cl | 7 | NPA169(1971)513 | 0.43 | | | 3 | |
| $^{36}$Cl | $^{35}$Cl(d,p)$^{36}$Cl | 12.3 | NPA481(1988)269 | | 0.68 | 0.68 | 1 | |
| $^{37}$Cl | $^{37}$Cl(p,d)$^{36}$Cl | 19 | NPA204(1973)609 | 30.1* | | | | AU |
| $^{37}$Cl | $^{37}$Cl(p,d)$^{36}$Cl | 35 | NPA239(1975)189 | | 1.58 | | 2 | |
| $^{37}$Cl | $^{37}$Cl(p,d)$^{36}$Cl | 40 | NPA189(1972)513 | | 0.66 | 0.97 | 4 | |
| $^{38}$Cl | $^{37}$Cl(d,p)$^{38}$Cl | 7.5 | NP83(1966)80 | 1.06* | | | 3 | BS |
| $^{38}$Cl | $^{37}$Cl(d,p)$^{38}$Cl | 12 | NPA225(1974)93 | | 1.81 | 1.81 | 3 | |
| $^{36}$Ar | $^{36}$Ar(p,d)$^{35}$Ar | 27.5 | NPA108(1968)113 | | 4.32 | | 5 | |
| $^{36}$Ar | $^{36}$Ar(p,d)$^{35}$Ar | 33.6 | PR172(1968)1078 | | 2.53 | 3.34 | 6 | |
| $^{37}$Ar | $^{36}$Ar(d,p)$^{37}$Ar | 9.162 | PRC3(1970)2314 | 0.29 | | | 6 | |
| $^{37}$Ar | $^{36}$Ar(d,p)$^{37}$Ar | 10.02 | PRC10(1974)1050 | | 0.34 | | 5 | |
| $^{37}$Ar | $^{36}$Ar(d,p)$^{37}$Ar | 18 | PRC4(1971)1778 | | 0.37 | 0.36 | 5 | |
| $^{38}$Ar | $^{38}$Ar(p,d)$^{37}$Ar | 26 | NPA250(1975)309 | | 2.47 | 2.47 | 6 | |
| $^{39}$Ar | $^{38}$Ar(d,p)$^{39}$Ar | 10.064 | PRC5(1972)1278 | | 0.87 | | 3 | |
| $^{39}$Ar | $^{38}$Ar(d,p)$^{39}$Ar | 11.6 | NPA114(1968)392 | | 0.77 | 0.813 | 4 | |
| $^{40}$Ar | $^{40}$Ar(p,d)$^{39}$Ar | 27.5 | NPA108(1968)113 | | 1.08 | 1.08 | 5 | |
| $^{40}$Ar | $^{40}$Ar(p,d)$^{39}$Ar | 35 | PRC16(1977)1357 | 2.25* | | | 4 | BS |
| $^{41}$Ar | $^{40}$Ar(d,p)$^{41}$Ar | 11.6 | NPA114(1968)392 | | 0.57 | | 2 | BS |
| $^{41}$Ar | $^{40}$Ar(d,p)$^{41}$Ar | 14.83 | NPA250(1975)45 | | 0.54 | 0.55 | 3 | |
| $^{39}$K | $^{39}$K(p,d)$^{38}$K | 35 | PRC10(1974)2184 | | 2.12 | | 4 | BS |
| $^{40}$K | $^{39}$K(d,p)$^{40}$K | 12 | NPA225(1974)93 | | 1.71 | 1.71 | 5 | |
| $^{41}$K | $^{41}$K(p,d)$^{40}$K | 15 | NPA213(1973)317 | | 0.91 | 0.91 | 3 | |
| $^{42}$K | $^{41}$K(d,p)$^{42}$K | 10 | NPA311(1978)61 | | 0.91 | | 1 | |
| $^{42}$K | $^{41}$K(d,p)$^{42}$K | 12 | NPA127(1969)343 | | 0.71 | 0.81 | 1 | |
| $^{40}$Ca | $^{40}$Ca(p,d)$^{39}$Ca | 27.3 | PL14(1965)113 | | 3.49 | | 3 | |
| $^{40}$Ca | $^{40}$Ca(p,d)$^{39}$Ca | 30 | NPA50(1964)49 | | 4.43 | | 4 | |
| $^{40}$Ca | $^{40}$Ca(p,d)$^{39}$Ca | 33.6 | PR172(1968)1078 | | 5.50 | | 3 | |
| $^{40}$Ca | $^{40}$Ca(p,d)$^{39}$Ca | 40 | NPA185(1972)465 | | 3.86 | | 3 | |
| $^{40}$Ca | $^{40}$Ca(p,d)$^{39}$Ca | 65 | PRC48(1993)95 | | 4.40 | 4.35 | 5 | |
| $^{40}$Ca | $^{40}$Ca(p,d)$^{39}$Ca | 65 | NPA343(1980)234 | 5.00* | | | 3 | NP |
| $^{41}$Ca | $^{40}$Ca(d,p)$^{41}$Ca | 4.13 | NP61(1965)209 | 1.36 | | | 1 | |
| $^{41}$Ca | $^{40}$Ca(d,p)$^{41}$Ca | 4.69 | NP61(1965)209 | 1.20 | | | 1 | |
| $^{41}$Ca | $^{40}$Ca(d,p)$^{41}$Ca | 5 | NPA109(1968)218 | 1.62 | | | 3 | |
| $^{41}$Ca | $^{40}$Ca(d,p)$^{41}$Ca | 5 | NPA172(1971)652 | 1.40 | | | 3 | |
| $^{41}$Ca | $^{40}$Ca(d,p)$^{41}$Ca | 6 | NPA109(1968)218 | 1.33 | | | 1 | |
| $^{41}$Ca | $^{40}$Ca(d,p)$^{41}$Ca | 6 | PRC14(1976)2082 | 1.24 | | | 2 | |
| $^{41}$Ca | $^{40}$Ca(d,p)$^{41}$Ca | 7 | NPA120(1968)401 | 1.25 | | | 3 | |
| $^{41}$Ca | $^{40}$Ca(d,p)$^{41}$Ca | 7 | PR136(1964)B971 | 1.00 | | | 1 | |
| $^{41}$Ca | $^{40}$Ca(d,p)$^{41}$Ca | 7.2 | NPA120(1968)401 | 1.27 | | | 3 | |
| $^{41}$Ca | $^{40}$Ca(d,p)$^{41}$Ca | 8 | PR136(1964)B971 | 1.17 | | | 3 | |
| $^{41}$Ca | $^{40}$Ca(d,p)$^{41}$Ca | 9 | NPA172(1971)652 | 1.05 | | | 5 | |
| $^{41}$Ca | $^{40}$Ca(d,p)$^{41}$Ca | 9 | PR136(1964)B971 | 1.19 | | | 3 | |
| $^{41}$Ca | $^{40}$Ca(d,p)$^{41}$Ca | 10 | NPA120(1968)421 | | 0.96 | | 3 | |
| $^{41}$Ca | $^{40}$Ca(d,p)$^{41}$Ca | 10 | NPA225(1974)267 | | 0.96 | | 1 | |
| $^{41}$Ca | $^{40}$Ca(d,p)$^{41}$Ca | 10 | PR136(1964)B971 | 1.07* | | | | BD |
| $^{41}$Ca | $^{40}$Ca(d,p)$^{41}$Ca | 11 | NP64(1965)241 | | 1.00 | | 3 | |
| $^{41}$Ca | $^{40}$Ca(d,p)$^{41}$Ca | 11 | NPA140(1970)577 | | | | | NP |
| $^{41}$Ca | $^{40}$Ca(d,p)$^{41}$Ca | 11 | NPA172(1971)652 | | 0.99 | | 4 | |
| $^{41}$Ca | $^{40}$Ca(d,p)$^{41}$Ca | 11 | NPA302(1978)12 | | 1.09 | | 4 | |
| $^{41}$Ca | $^{40}$Ca(d,p)$^{41}$Ca | 11 | PR136(1964)B971 | 1.43* | | | 3 | BD |
| $^{41}$Ca | $^{40}$Ca(d,p)$^{41}$Ca | 11 | PR181(1969)1529 | | 0.98 | | 3 | |
| $^{41}$Ca | $^{40}$Ca(d,p)$^{41}$Ca | 11 | PRC14(1976)946 | | 1.02 | | 2 | |
| $^{41}$Ca | $^{40}$Ca(d,p)$^{41}$Ca | 11.8 | NP53(1964)77 | | 0.99 | | 1 | |
| $^{41}$Ca | $^{40}$Ca(d,p)$^{41}$Ca | 12 | NPA140(1970)577 | | 0.99 | | 2 | |
| $^{41}$Ca | $^{40}$Ca(d,p)$^{41}$Ca | 12 | NPA243(1975)100 | | 1.07 | | 2 | |
| $^{41}$Ca | $^{40}$Ca(d,p)$^{41}$Ca | 12 | PR136(1964)B971 | 1.04* | | | 3 | BS |
| $^{41}$Ca | $^{40}$Ca(d,p)$^{41}$Ca | 12.8 | PR146(1966)799 | | 1.11 | | 1 | |
| $^{41}$Ca | $^{40}$Ca(d,p)$^{41}$Ca | 14.3 | PR138(1965)B1425 | | 1.00 | | 5 | |
| $^{41}$Ca | $^{40}$Ca(d,p)$^{41}$Ca | 20 | NPA506(1990)159 | | 1.04 | 1.01 | 2 | |
| $^{41}$Ca | $^{40}$Ca(d,p)$^{41}$Ca | 56 | NPA419(1984)530 | 0.76* | | | 4 | BS |
| $^{41}$Ca | $^{40}$Ca(d,p)$^{41}$Ca | 56 | PRC50(1994)263 | 1.07* | | | 3 | BS |
| $^{42}$Ca | $^{41}$Ca(d,p)$^{42}$Ca | 11 | NPA302(1978)12 | | 1.92 | | 2 | |
| $^{42}$Ca | $^{41}$Ca(d,p)$^{42}$Ca | 12 | NPA243(1975)100 | | 1.78 | | 5 | |
| $^{42}$Ca | $^{41}$Ca(d,p)$^{42}$Ca | 12 | PLB40(1972)641 | | 1.81 | 1.82 | 3 | |
| $^{42}$Ca | $^{42}$Ca(p,d)$^{41}$Ca | 26.5 | NPA113(1968)303 | | 2.18 | | 4 | |
| $^{42}$Ca | $^{42}$Ca(p,d)$^{41}$Ca | 40 | NPA185(1972)465 | | 2.00 | 2.12 | 2 | |
| $^{43}$Ca | $^{42}$Ca(d,p)$^{43}$Ca | 7 | NPA120(1968)401 | 0.85 | | | 3 | |
| $^{43}$Ca | $^{42}$Ca(d,p)$^{43}$Ca | 7.2 | NPA120(1968)401 | 0.93 | | | 3 | |
| $^{43}$Ca | $^{42}$Ca(d,p)$^{43}$Ca | 7.2 | PR146(1966)734 | 0.84 | | | 3 | |
| $^{43}$Ca | $^{42}$Ca(d,p)$^{43}$Ca | 10 | NPA120(1968)421 | | 0.66 | | 2 | |
| $^{43}$Ca | $^{42}$Ca(d,p)$^{43}$Ca | 10 | NPA225(1974)267 | | 0.59 | 0.63 | 2 | |
| $^{43}$Ca | $^{43}$Ca(p,d)$^{42}$Ca | 40 | PRC7(1973)637 | | 0.63 | 0.63 | 3 | |
| $^{44}$Ca | $^{43}$Ca(d,p)$^{44}$Ca | 8.5 | PR155(1967)1229 | 5.14 | | 5.14 | 3 | |
| $^{44}$Ca | $^{44}$Ca(p,d)$^{43}$Ca | 17.5 | PR144(1966)941 | | 2.84 | | 2 | |
| $^{44}$Ca | $^{44}$Ca(p,d)$^{43}$Ca | 26.5 | NPA113(1968)303 | | 5.34 | | 4 | |
| $^{44}$Ca | $^{44}$Ca(p,d)$^{43}$Ca | 40 | NPA185(1972)465 | | 3.23 | 3.93 | 5 | |
| $^{45}$Ca | $^{44}$Ca(d,p)$^{45}$Ca | 7 | NPA120(1968)401 | 0.55 | | | 3 | |
| $^{45}$Ca | $^{44}$Ca(d,p)$^{45}$Ca | 7 | PR156(1967)1255 | 0.62 | | | 2 | |
| $^{45}$Ca | $^{44}$Ca(d,p)$^{45}$Ca | 7.2 | NPA120(1968)401 | 0.54 | | | 2 | |
| $^{45}$Ca | $^{44}$Ca(d,p)$^{45}$Ca | 10 | NPA120(1968)421 | | 0.37 | | 2 | |
| $^{45}$Ca | $^{44}$Ca(d,p)$^{45}$Ca | 10 | NPA225(1974)267 | | 0.37 | 0.37 | 2 | |
| $^{47}$Ca | $^{46}$Ca(d,p)$^{47}$Ca | 7 | NPA120(1968)401 | 0.35 | | | 3 | |
| $^{47}$Ca | $^{46}$Ca(d,p)$^{47}$Ca | 7.2 | NPA120(1968)401 | 0.29 | | | 3 | |
| $^{47}$Ca | $^{46}$Ca(d,p)$^{47}$Ca | 10 | NPA120(1968)421 | | 0.26 | | 2 | |
| $^{47}$Ca | $^{46}$Ca(d,p)$^{47}$Ca | 10 | PR138(1965)B1097 | | 0.26 | 0.26 | 4 | |
| $^{48}$Ca | $^{48}$Ca(p,d)$^{47}$Ca | 17.5 | PR144(1966)941 | | 8.82 | | 5 | |
| $^{48}$Ca | $^{48}$Ca(p,d)$^{47}$Ca | 18 | PR170(1968)1003 | | 5.51 | | 4 | |
| $^{48}$Ca | $^{48}$Ca(p,d)$^{47}$Ca | 40 | NPA185(1972)465 | | 7.35 | 7.35 | 3 | |
| $^{49}$Ca | $^{48}$Ca(d,p)$^{49}$Ca | 4.5 | NPA160(1971)289 | 0.77 | | | 4 | |
| $^{49}$Ca | $^{48}$Ca(d,p)$^{49}$Ca | 5 | NPA160(1971)289 | 0.76 | | | 3 | |
| $^{49}$Ca | $^{48}$Ca(d,p)$^{49}$Ca | 5.5 | NPA160(1971)289 | 0.73 | | | 3 | |
| $^{49}$Ca | $^{48}$Ca(d,p)$^{49}$Ca | 7 | NPA120(1968)401 | 0.81 | | | 3 | |
| $^{49}$Ca | $^{48}$Ca(d,p)$^{49}$Ca | 7 | NPA160(1971)289 | 0.89 | | | 4 | |
| $^{49}$Ca | $^{48}$Ca(d,p)$^{49}$Ca | 7 | PR135(1964)B865 | 1.5 | | | 4 | |
| $^{49}$Ca | $^{48}$Ca(d,p)$^{49}$Ca | 7.2 | NPA120(1968)401 | 0.87 | | | 3 | |
| $^{49}$Ca | $^{48}$Ca(d,p)$^{49}$Ca | 10 | NPA120(1968)421 | 0.79* | | | 1 | NP |
| $^{49}$Ca | $^{48}$Ca(d,p)$^{49}$Ca | 10 | NPA225(1974)267 | | 0.63 | | 2 | |
| $^{49}$Ca | $^{48}$Ca(d,p)$^{49}$Ca | 11.9 | NPA303(1978)121 | 0.61* | | | 2 | NP |
| $^{49}$Ca | $^{48}$Ca(d,p)$^{49}$Ca | 13 | PRC12(1975)827 | | 0.77 | | 3 | |
| $^{49}$Ca | $^{48}$Ca(d,p)$^{49}$Ca | 16 | PRC12(1975)827 | | 0.68 | | 3 | |
| $^{49}$Ca | $^{48}$Ca(d,p)$^{49}$Ca | 19.3 | PRC12(1975)827 | | 0.64 | 0.69 | 1 | |
| $^{49}$Ca | $^{48}$Ca(d,p)$^{49}$Ca | 56 | NPA576(1994)123 | 0.66* | | | 3 | BS |
| $^{45}$Sc | $^{45}$Sc(p,d)$^{44}$Sc | 17.5 | PR134(1964)B378 | | 0.30 | 0.30 | 3 | BS |
| $^{46}$Sc | $^{45}$Sc(d,p)$^{46}$Sc | 7 | PR151(1966)939. | 0.39 | | | 2 | |
| $^{46}$Sc | $^{45}$Sc(d,p)$^{46}$Sc | 12 | PRC46(1992)144 | | 0.51 | 0.51 | 2 | |
| $^{46}$Ti | $^{46}$Ti(p,d)$^{45}$Ti | 17.5 | PR135(1964)B389 | | 2.6 | | 3 | |
| $^{46}$Ti | $^{46}$Ti(p,d)$^{45}$Ti | 26 | NPA111(1968)449 | | 2.29 | 2.423 | 4 | |
| $^{46}$Ti | $^{46}$Ti(p,d)$^{45}$Ti | 34.78 | NPA152(1970)609 | 1.28* | | | 3 | |
| $^{47}$Ti | $^{46}$Ti(d,p)$^{47}$Ti | 7 | NP73(1965)321 | | 0.03 | 0.03 | 4 | BS |
| $^{47}$Ti | $^{46}$Ti(d,p)$^{47}$Ti | 10 | NPA196(1972)225 | 0.01* | | | 4 | BD |
| $^{48}$Ti | $^{47}$Ti(d,p)$^{48}$Ti | 13.6 | J,YF,25,16,77 | | 0.14 | 0.14 | 1 | BS |
| $^{48}$Ti | $^{48}$Ti(p,d)$^{47}$Ti | 24.8 | NPA152(1970)609 | 0.10* | | | 4 | BD |
| $^{48}$Ti | $^{48}$Ti(p,d)$^{47}$Ti | 29.82 | NPA152(1970)609 | 0.12* | | | 3 | BD |
| $^{48}$Ti | $^{48}$Ti(p,d)$^{47}$Ti | 35.15 | NPA152(1970)609 | | 0.11 | | 3 | |
| $^{48}$Ti | $^{48}$Ti(p,d)$^{47}$Ti | 39.97 | NPA152(1970)609 | | 0.11 | | 3 | |
| $^{48}$Ti | $^{48}$Ti(p,d)$^{47}$Ti | 45.05 | NPA152(1970)609 | | 0.097 | 0.11 | 3 | |
| $^{49}$Ti | $^{48}$Ti(d,p)$^{49}$Ti | 6 | PR159(1967)920 | 0.3 | | | 4 | |
| $^{49}$Ti | $^{48}$Ti(d,p)$^{49}$Ti | 21.4 | PR131(1963)811 | | 0.23 | 0.23 | 3 | |
| $^{49}$Ti | $^{49}$Ti(p,d)$^{48}$Ti | 17.5 | PR135(1964)B389 | | 0.25 | | 4 | |
| $^{49}$Ti | $^{49}$Ti(p,d)$^{48}$Ti | 20.9 | NPA177(1971)205 | | 0.27 | 0.26 | 4 | |
| $^{50}$Ti | $^{49}$Ti(d,p)$^{50}$Ti | 13.6 | J,YF,25,16,77 | | 6.23 | | 4 | |
| $^{50}$Ti | $^{49}$Ti(d,p)$^{50}$Ti | 21.4 | PR131(1963)811 | | 8 | 7.115 | 4 | |
| $^{50}$Ti | $^{50}$Ti(p,d)$^{49}$Ti | 17.5 | PR135(1964)B389 | | 5.98 | | 4 | |
| $^{50}$Ti | $^{50}$Ti(p,d)$^{49}$Ti | 45.05 | NPA152(1970)609 | | 4.86 | 5.50 | 3 | |
| $^{51}$Ti | $^{50}$Ti(d,p)$^{51}$Ti | 6 | PR136(1964)B438 | 0.53* | | | 3 | |
| $^{51}$Ti | $^{50}$Ti(d,p)$^{51}$Ti | 21.4 | PR131(1963)811 | | 1.25 | 1.25 | 5 | |
| $^{51}$V | $^{50}$V(d,p)$^{51}$V | 7.5 | NPA94(1967)673 | | 1.58 | 1.58 | 3 | |
| $^{51}$V | $^{51}$V(p,d)$^{50}$V | 18.5 | PR134(1964)B752 | | 1.33 | | 3 | BS |
| $^{51}$V | $^{51}$V(p,d)$^{50}$V | 51.9 | PLB73(1978)145 | | 0.75 | 1.098 | 2 | BS |
| $^{50}$Cr | $^{50}$Cr(p,d)$^{49}$Cr | 17.5 | PR160(1967)997 | 0.11* | | | 5 | BS |
| $^{50}$Cr | $^{50}$Cr(p,d)$^{49}$Cr | 55 | NPA435(1985)7 | | 0.11 | 0.11 | 3 | BS |
| $^{51}$Cr | $^{50}$Cr(d,p)$^{51}$Cr | 6.6 | NP51(1964)161 | 0.62 | | | 2 | |
| $^{51}$Cr | $^{50}$Cr(d,p)$^{51}$Cr | 7.5 | PR170(1968)1013 | 0.67 | | | 2 | |
| $^{51}$Cr | $^{50}$Cr(d,p)$^{51}$Cr | 10 | NPA198(1972)237 | 2.83* | | | 3 | AU |
| $^{51}$Cr | $^{50}$Cr(d,p)$^{51}$Cr | 12 | NPA282(1977)87 | | 0.30 | 0.30 | 3 | |
| $^{52}$Cr | $^{52}$Cr(p,d)$^{51}$Cr | 17.5 | PR160(1967)997 | | 6.55 | | 6 | |
| $^{52}$Cr | $^{52}$Cr(p,d)$^{51}$Cr | 18.5 | PR134(1964)B752 | | 5.87 | 6.24 | 5 | |
| $^{53}$Cr | $^{52}$Cr(d,p)$^{53}$Cr | 5.41 | NP84(1966)398 | 0.67 | | | 3 | |
| $^{53}$Cr | $^{52}$Cr(d,p)$^{53}$Cr | 5.72 | NP84(1966)398 | 0.57 | | | 4 | |
| $^{53}$Cr | $^{52}$Cr(d,p)$^{53}$Cr | 6 | NPA277(1977)374 | 0.46 | | | 4 | |
| $^{53}$Cr | $^{52}$Cr(d,p)$^{53}$Cr | 6.02 | NP84(1966)398 | 0.53 | | | 2 | |
| $^{53}$Cr | $^{52}$Cr(d,p)$^{53}$Cr | 6.33 | NP84(1966)398 | 0.49 | | | 3 | |
| $^{53}$Cr | $^{52}$Cr(d,p)$^{53}$Cr | 7.5 | NPA121(1968)1 | 0.54 | | | 3 | |
| $^{53}$Cr | $^{52}$Cr(d,p)$^{53}$Cr | 9.14 | NP86(1966)65 | 0.36 | | | 3 | |
| $^{53}$Cr | $^{52}$Cr(d,p)$^{53}$Cr | 10 | NPA196(1972)225 | | 0.43 | | 2 | |
| $^{53}$Cr | $^{52}$Cr(d,p)$^{53}$Cr | 10 | NPA206(1973)225 | | 0.42 | | 2 | |
| $^{53}$Cr | $^{52}$Cr(d,p)$^{53}$Cr | 10 | NPA277(1977)119 | | 0.39 | | 1 | |
| $^{53}$Cr | $^{52}$Cr(d,p)$^{53}$Cr | 10 | NP72(1965)273 | | 0.33 | | 1 | BD |
| $^{53}$Cr | $^{52}$Cr(d,p)$^{53}$Cr | 10.15 | NP86(1966)65 | | 0.37 | | 3 | |
| $^{53}$Cr | $^{52}$Cr(d,p)$^{53}$Cr | 11.18 | NP86(1966)65 | | 0.36 | | 3 | |
| $^{53}$Cr | $^{52}$Cr(d,p)$^{53}$Cr | 12 | NPA167(1971)289 | | 0.42 | | 4 | |
| $^{53}$Cr | $^{52}$Cr(d,p)$^{53}$Cr | 22 | NPA573(1994)1 | | 0.36 | 0.39 | 2 | |
| $^{53}$Cr | $^{53}$Cr(p,d)$^{52}$Cr | 16.6 | NPA177(1971)205 | | 0.37 | 0.37 | 2 | |
| $^{55}$Cr | $^{54}$Cr(d,p)$^{55}$Cr | 8 | NPA142(1970)469 | | 0.63 | | 2 | |
| $^{55}Cr$ | $^{54}Cr(d,p)^{55}Cr$ | 10 | NPA337(1980)389 | 1.1* | | | 2 | NP |
| $^{55}Cr$ | $^{54}Cr(d,p)^{55}Cr$ | 10 | NP72(1965)273 | 0.87* | | 0.63 | 3 | BD |

Table II List of nuclei with spectroscopic factors obtained from both (p,d) and (d,p) reactions. $N_{pd}$ and $N_{dp}$ denote the number of (p,d) and (d,p) independent measurements included in the analysis.

| A | A(p,d)B | $N_{pd}$ | B(d,p)A | $N_{dp}$ |
|---|---|---|---|---|
| $^{11}$Be | 0.57 | 1 | 0.49 | 2 |
| $^{11}$B | 1.29 | 1 | 1.55 | 3 |
| $^{13}$C | 0.81 | 4 | 0.73 | 12 |
| $^{14}$C | 1.50 | 3 | 1.82 | 2 |
| $^{15}$N | 1.65 | 2 | 1.39 | 4 |
| $^{17}$O | 0.81 | 3 | 0.98 | 10 |
| $^{18}$O | 1.60 | 2 | 1.80 | 1 |
| $^{21}$Ne | 0.03 | 1 | 0.04 | 2 |
| $^{26}$Mg | 2.80 | 3 | 2.01 | 1 |
| $^{30}$Si | 0.87 | 1 | 0.79 | 2 |
| $^{42}$Ca | 2.12 | 2 | 1.82 | 3 |
| $^{43}$Ca | 0.63 | 1 | 0.63 | 2 |
| $^{44}$Ca | 3.93 | 3 | 5.14 | 1 |
| $^{48}$Ti | 0.11 | 3 | 0.14 | 1 |
| $^{49}$Ti | 0.26 | 2 | 0.23 | 1 |
| $^{50}$Ti | 5.50 | 2 | 7.12 | 2 |
| $^{51}$V | 1.10 | 2 | 1.58 | 1 |
| $^{53}$Cr | 0.37 | 1 | 0.39 | 8 |

Table III: List of isotopes with the extracted spectroscopic factors and other information such as the mass number (A), charge number (Z) and neutron number (N) for the nuclei. $j^{\pi}$, T and $S_n$ are the spin and parity, isospin and neutron separation energy of the nuclei. For completeness, we also list the root mean square radii of the neutron wave-functions. As the spectroscopic factors obtained in the present work come from using conventional parameters in the TWOFNR [13] calculations, we label these values as SF(conv). Endt's compiled values are also listed when available. The model space and interactions used in Oxbash [251] are listed together with the predicted SF values labeled as LB-SM.

| Isotope | A | Z | N | $j^{\pi}$ | T | S_n | rms | Endt | SF (conv) | LB-SM | Model Space | Interaction |
|---|---|---|---|---|---|---|---|---|---|---|---|---|
| $^{6}$Li | 6 | 3 | 3 | $1/2^-$ | 0 | 5.66 | 2.91 | | 1.12 ± 0.32 | 0.68 | PPN | CKPPN |
| $^{7}$Li | 7 | 3 | 4 | $1/2^-$ | 1/2 | 7.25 | 2.81 | | 1.85 ± 0.37 | 0.63 | PPN | CKPPN |
| $^{8}$Li | 8 | 3 | 5 | $1/2^-$ | 1 | 2.03 | 3.66 | | 0.62 ± 0.18 | 1.09 | PPN | CKPPN |
| $^{9}$Li | 9 | 3 | 6 | $1/2^-$ | 3/2 | 4.06 | 3.23 | | 0.98 ± 0.28 | 0.81 | PPN | CKPPN |
| $^{9}$Be | 9 | 4 | 5 | $3/2^-$ | 1/2 | 1.67 | 3.86 | | 0.45 ± 0.03 | 0.57 | PPN | CKPPN |
| $^{10}$Be | 10 | 4 | 6 | $3/2^-$ | 1 | 6.81 | 2.96 | | 1.58 ± 0.15 | 2.36 | PPN | CKPPN |
| $^{11}$Be | 11 | 4 | 7 | $1/2^+$ | 3/2 | 0.50 | 7.11 | | 0.51 ± 0.06 | 0.74 | SPSDPF | WBP |
| $^{10}$B | 10 | 5 | 5 | $3/2^-$ | 0 | 8.44 | 2.85 | | 0.50 ± 0.07 | 0.60 | PPN | CKPPN |
| $^{11}$B | 11 | 5 | 6 | $3/2^-$ | 1/2 | 11.45 | 2.73 | | 1.48 ± 0.19 | 1.09 | PPN | CKPPN |
| $^{12}$B | 12 | 5 | 7 | $1/2^-$ | 1 | 3.37 | 3.46 | | 0.45 ± 0.06 | 0.83 | PPN | CKPPN |
| $^{12}$C | 12 | 6 | 6 | $3/2^-$ | 0 | 18.72 | 2.53 | | 3.12 ± 0.36 | 2.85 | PPN | CKPPN |
| $^{13}$C | 13 | 6 | 7 | $1/2^-$ | 1/2 | 4.95 | 3.26 | | 0.75 ± 0.10 | 0.61 | PPN | CKPPN |
| $^{14}$C | 14 | 6 | 8 | $1/2^-$ | 1 | 8.18 | 3.00 | | 1.63 ± 0.33 | 1.73 | PPN | CKPPN |
| $^{15}$C | 15 | 6 | 9 | $1/2^+$ | 3/2 | 1.22 | 5.51 | | 1.12 ± 0.32 | 0.98 | SPSDPF | WBP |
| $^{14}$N | 14 | 7 | 7 | $1/2^-$ | 0 | 10.55 | 2.87 | | 0.77 ± 0.12 | 0.69 | PPN | CKPPN |
| $^{15}$N | 15 | 7 | 8 | $1/2^-$ | 1/2 | 10.83 | 2.89 | | 1.48 ± 0.24 | 1.46 | PPN | CKPPN |
| $^{16}$N | 16 | 7 | 9 | $3/2^+$ | 1 | 2.49 | 4.26 | | 0.42 ± 0.12 | 0.96 | SPSDPF | WBP |
| $^{16}$O | 16 | 8 | 8 | $1/2^-$ | 0 | 15.66 | 2.74 | | 2.46 ± 0.26 | 2.00 | PPN | CKPPN |
| $^{17}$O | 17 | 8 | 9 | $5/2^+$ | 1/2 | 4.14 | 3.48 | | 0.94 ± 0.13 | 1.00 | SDPN | WPN |
| $^{18}$O | 18 | 8 | 10 | $5/2^+$ | 1 | 8.04 | 3.24 | | 1.66 ± 0.19 | 1.58 | SDPN | WPN |
| $^{19}$O | 19 | 8 | 11 | $5/2^+$ | 3/2 | 3.95 | 3.57 | | 0.43 ± 0.06 | 0.69 | SDPN | WPN |
| $^{19}$F | 19 | 9 | 10 | $1/2^+$ | 1/2 | 10.43 | 2.66 | | 1.60 ± 0.23 | 0.56 | SDPN | WPN |
| $^{20}$F | 20 | 9 | 11 | $3/2^+$ | 1 | 6.60 | 3.39 | | ~0.01 | 0.02 | SDPN | WPN |
| $^{21}$Ne | 21 | 10 | 11 | $3/2^+$ | 1/2 | 6.76 | 3.41 | 0.01 | 0.03 ± 0.01 | 0.03 | SD | W |
| $^{22}$Ne | 22 | 10 | 12 | $3/2^+$ | 1 | 10.36 | 3.27 | 0.19 | 0.24 ± 0.03 | 0.01 | SDPN | WPN |
| $^{23}$Ne | 23 | 10 | 13 | $5/2^+$ | 3/2 | 5.20 | 3.58 | 0.24 | 0.24 ± 0.03 | 0.03 | SDPN | WPN |
| $^{24}$Na | 24 | 11 | 13 | $1/2^+$ | 1 | 8.87 | 3.49 | 0.30 | 0.59 ± 0.17 | 0.39 | SDPN | WPN |
| $^{24}$Mg | 24 | 12 | 12 | $3/2^+$ | 0 | 16.53 | 3.13 | | 0.41 ± 0.06 | 0.22 | SDPN | WPN |
| $^{25}$Mg | 25 | 12 | 13 | $5/2^+$ | 1/2 | 7.33 | 3.50 | 0.37 | 0.29 ± 0.03 | 0.34 | SDPN | WPN |
| $^{26}$Mg | 26 | 12 | 14 | $5/2^+$ | 1 | 11.09 | 3.35 | 1.80 | 2.43 ± 0.50 | 2.51 | SDPN | WPN |
| $^{27}$Mg | 27 | 12 | 15 | $1/2^+$ | 3/2 | 6.44 | 3.90 | 0.58 | 0.45 ± 0.13 | 0.46 | SDPN | WPN |
| $^{27}$Al | 27 | 13 | 14 | $5/2^+$ | 1/2 | 13.06 | 3.31 | 1.10 | 1.40 ± 0.20 | 1.10 | SDPN | WPN |
| $^{28}$Al | 28 | 13 | 15 | $1/2^+$ | 1 | 7.73 | 3.78 | 0.50 | 0.66 ± 0.10 | 0.60 | SDPN | WPN |
| $^{28}$Si | 28 | 14 | 14 | $5/2^+$ | 0 | 17.18 | 3.22 | | 4.40 ± 1.24 | 3.62 | SDPN | WPN |
| $^{29}$Si | 29 | 14 | 15 | 1/2$^+$ | 1/2 | 8.47 | 3.73 | 0.55 | 0.42 $\pm$ 0.13 | 0.45 | SDPN | WPN |
| $^{30}$Si | 30 | 14 | 16 | 1/2$^+$ | 1 | 10.61 | 2.87 | 0.89 | 0.84 $\pm$ 0.10 | 0.82 | SDPN | WPN |
| $^{31}$Si | 31 | 14 | 17 | 3/2$^+$ | 3/2 | 6.59 | 3.70 | 0.75 | 0.54 $\pm$ 0.09 | 0.58 | SDPN | WPN |
| $^{32}$P | 32 | 15 | 17 | 1/2$^+$ | 1 | 7.94 | 3.64 | 0.80 | 0.58 $\pm$ 0.10 | 0.60 | SDPN | WPN |
| $^{32}$S | 32 | 16 | 16 | 1/2$^+$ | 0 | 15.04 | 3.40 | | 1.51 $\pm$ 0.43 | 0.96 | SDPN | WPN |
| $^{33}$S | 33 | 16 | 17 | 3/2$^+$ | 1/2 | 8.64 | 3.63 | 0.70 | 0.70 $\pm$ 0.20 | 0.61 | SDPN | WPN |
| $^{34}$S | 34 | 16 | 18 | 3/2$^+$ | 1 | 11.42 | 3.53 | 1.90 | 1.43 $\pm$ 0.35 | 1.83 | SDPN | WPN |
| $^{35}$S | 35 | 16 | 19 | 3/2$^+$ | 3/2 | 6.99 | 3.77 | 0.38 | 0.30 $\pm$ 0.09 | 0.36 | SDPN | WPN |
| $^{37}$S | 37 | 16 | 21 | 7/2$^-$ | 5/2 | 4.30 | 4.02 | | 0.88 $\pm$ 0.12 | | | |
| $^{35}$Cl | 35 | 17 | 18 | 3/2$^+$ | 1/2 | 12.64 | 3.51 | | 0.35 $\pm$ 0.10 | 0.32 | SDPN | WPN |
| $^{36}$Cl | 36 | 17 | 19 | 1/2$^+$ | 1 | 8.58 | 3.70 | 1.20 | 0.68 $\pm$ 0.19 | 0.77 | SDPN | WPN |
| $^{37}$Cl | 37 | 17 | 20 | 1/2$^+$ | 3/2 | 10.31 | 3.64 | 0.95 | 0.97 $\pm$ 0.43 | 1.15 | SDPN | WPN |
| $^{38}$Cl | 38 | 17 | 21 | 1/2$^-$ | 2 | 6.11 | 3.94 | 0.78 | 1.81 $\pm$ 0.51 | | | |
| $^{36}$Ar | 36 | 18 | 18 | 3/2$^+$ | 0 | 15.26 | 3.45 | | 3.34 $\pm$ 0.89 | 2.06 | SDPN | WPN |
| $^{37}$Ar | 37 | 18 | 19 | 3/2$^+$ | 1/2 | 8.79 | 3.71 | 0.49 | 0.36 $\pm$ 0.05 | 0.36 | SDPN | WPN |
| $^{38}$Ar | 38 | 18 | 20 | 3/2$^+$ | 1 | 11.84 | 3.60 | 2.50 | 2.47 $\pm$ 0.70 | 3.04 | SDPN | WPN |
| $^{39}$Ar | 39 | 18 | 21 | 7/2$^-$ | 3/2 | 6.60 | 3.94 | 0.64 | 0.81 $\pm$ 0.11 | | | |
| $^{40}$Ar | 40 | 18 | 22 | 7/2$^-$ | 2 | 9.87 | 3.83 | 1.20 | 1.08 $\pm$ 0.31 | | | |
| $^{41}$Ar | 41 | 18 | 23 | 7/2$^-$ | 5/2 | 6.10 | 4.01 | 0.47 | 0.55 $\pm$ 0.08 | | | |
| $^{39}$K | 39 | 19 | 20 | 3/2$^+$ | 1/2 | 13.08 | 3.58 | 2.00 | 2.12 $\pm$ 0.60 | 1.72 | SDPN | WPN |
| $^{40}$K | 40 | 19 | 21 | 5/2$^-$ | 1 | 7.80 | 3.90 | 0.94 | 1.71 $\pm$ 0.48 | | | |
| $^{41}$K | 41 | 19 | 22 | 5/2$^-$ | 3/2 | 10.10 | 3.84 | 0.56 | 0.91 $\pm$ 0.26 | | | |
| $^{42}$K | 42 | 19 | 23 | 1/2$^-$ | 2 | 7.53 | 3.96 | 0.34 | 0.81 $\pm$ 0.11 | | | |
| $^{40}$Ca | 40 | 20 | 20 | 3/2$^+$ | 0 | 15.64 | 3.81 | | 4.35 $\pm$ 0.62 | 4.00 | SDPN | WPN |
| $^{41}$Ca | 41 | 20 | 21 | 7/2$^-$ | 1/2 | 8.36 | 3.90 | 0.85 | 1.01 $\pm$ 0.06 | 1.00 | FPPN | FPBPPN |
| $^{42}$Ca | 42 | 20 | 22 | 7/2$^-$ | 1 | 11.48 | 3.82 | 1.60 | 1.93 $\pm$ 0.17 | 1.81 | FPPN | FPBPPN |
| $^{43}$Ca | 43 | 20 | 23 | 7/2$^-$ | 3/2 | 7.93 | 3.97 | 0.58 | 0.63 $\pm$ 0.07 | 0.75 | FPPN | FPBPPN |
| $^{44}$Ca | 44 | 20 | 24 | 7/2$^-$ | 2 | 11.13 | 3.87 | 3.10 | 3.93 $\pm$ 1.08 | 3.64 | FPPN | FPBPPN |
| $^{45}$Ca | 45 | 20 | 25 | 7/2$^-$ | 5/2 | 7.41 | 4.03 | | 0.37 $\pm$ 0.05 | 0.50 | FPPN | FPBPPN |
| $^{47}$Ca | 47 | 20 | 27 | 7/2$^-$ | 7/2 | 7.28 | 4.08 | | 0.26 $\pm$ 0.04 | 0.26 | FPPN | FPBPPN |
| $^{48}$Ca | 48 | 20 | 28 | 7/2$^-$ | 4 | 9.95 | 3.99 | | 7.35 $\pm$ 1.42 | 7.38 | FPPN | FPBPPN |
| $^{49}$Ca | 49 | 20 | 29 | 3/2$^-$ | 9/2 | 5.15 | 4.59 | | 0.69 $\pm$ 0.07 | 0.92 | FPPN | FPBPPN |
| $^{45}$Sc | 45 | 21 | 24 | 3/2$^-$ | 3/2 | 11.32 | 3.89 | 0.34 | 0.30 $\pm$ 0.08 | 0.35 | FPPN | FPBPPN |
| $^{46}$Sc | 46 | 21 | 25 | 1/2$^-$ | 2 | 8.76 | 4.00 | | 0.51 $\pm$ 0.14 | 0.37 | FPPN | FPBPPN |
| $^{46}$Ti | 46 | 22 | 24 | 7/2$^-$ | 1 | 13.19 | 3.85 | | 2.42 $\pm$ 0.34 | 2.58 | FPPN | FPBPPN |
| $^{47}$Ti | 47 | 22 | 25 | 5/2$^-$ | 3/2 | 8.88 | 4.01 | | ~0.03 | | | |
| $^{48}$Ti | 48 | 22 | 26 | 5/2$^-$ | 2 | 11.63 | 3.94 | | 0.11 $\pm$ 0.01 | | | |
| $^{49}$Ti | 49 | 22 | 27 | 7/2$^-$ | 5/2 | 8.14 | 4.08 | | 0.25 $\pm$ 0.03 | | | |
| $^{50}$Ti | 50 | 22 | 28 | 7/2$^-$ | 3 | 10.94 | 4.00 | | 6.36 $\pm$ 1.10 | | | |
| $^{51}$Ti | 51 | 22 | 29 | 3/2$^-$ | 7/2 | 6.37 | 4.46 | | 1.25 $\pm$ 0.35 | | | |
| $^{51}$V | 51 | 23 | 28 | 5/2$^-$ | 5/2 | 11.05 | 4.01 | | 1.28 $\pm$ 0.32 | | | |
| $^{50}$Cr | 50 | 24 | 26 | 5/2$^-$ | 1 | 13.00 | 3.94 | | 0.11 $\pm$ 0.02 | | | |
| $^{51}$Cr | 51 | 24 | 27 | 7/2$^-$ | 3/2 | 9.26 | 4.08 | | 0.30 $\pm$ 0.08 | | | |
| $^{52}$Cr | 52 | 24 | 28 | 7/2$^-$ | 2 | 12.04 | 4.00 | | 6.24 $\pm$ 0.88 | | | |
| $^{53}$Cr | 53 | 24 | 29 | 3/2$^-$ | 5/2 | 7.94 | 4.34 | | 0.39 $\pm$ 0.03 | | | |
| $^{55}$Cr | 55 | 24 | 31 | 3/2$^-$ | 7 /2 | 6.24 | 4.03 | | 0.63 $\pm$ 0.13 | | | |

Figure 1: (Color online) Comparison of tabulated data (closed points) [21] and digitized data (open points) [82] from the same measurement of the angular distributions of the protons obtained in the $^{14}N(d,p)^{15}N$ reaction at incident deuteron energy of 12 MeV. The curve is the predicted angular distributions from the code TWOFNR as described in the text, multiplied by 1.12 which is the spectroscopic factor.

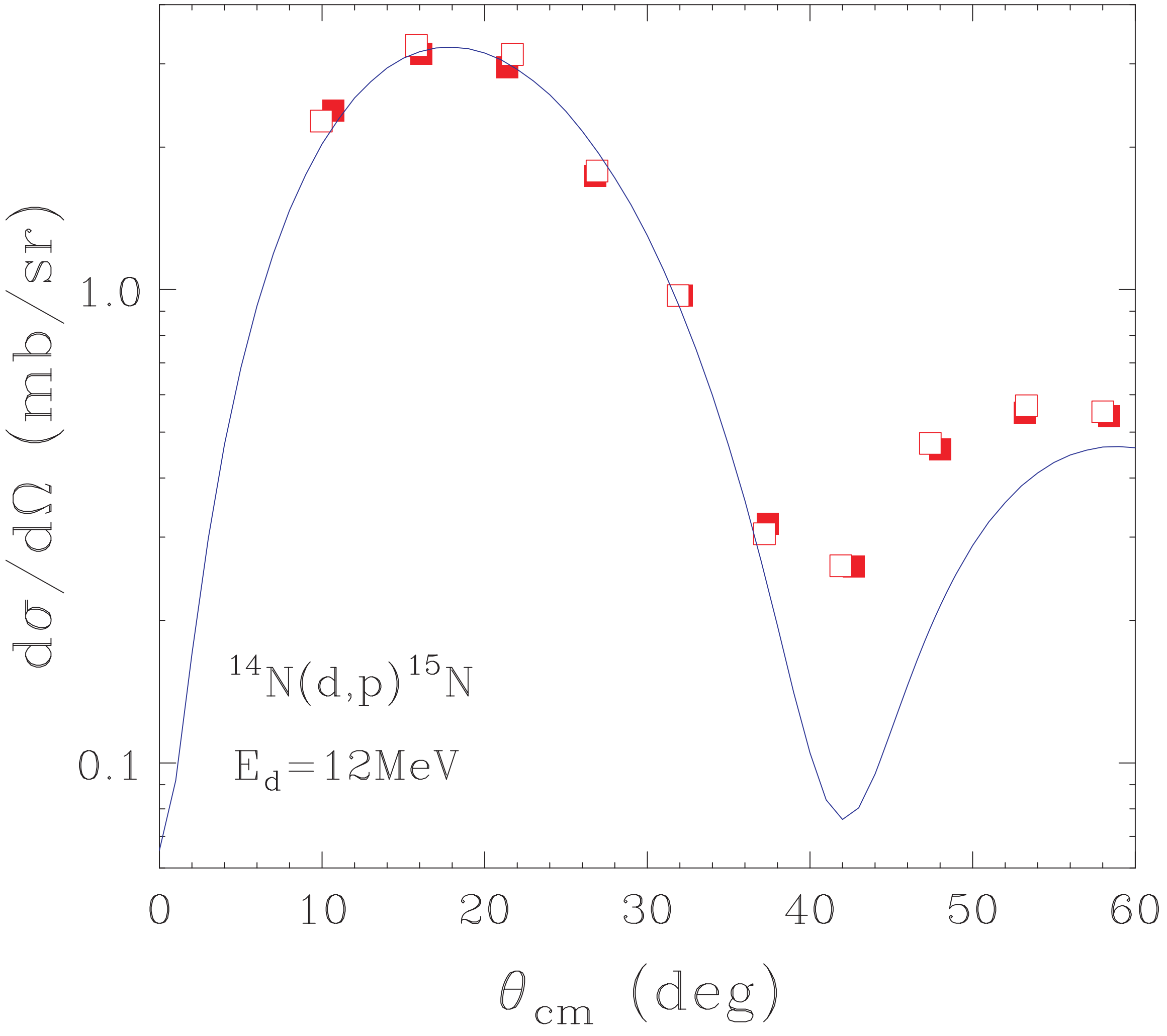

Figure 2: (Color online) The angular distributions of the deuteron obtained in the $^{44}$Ca(p,d) $^{43}$Ca reaction at incident proton energy of 40 MeV [174]. The curve is the predicted angular distributions from the code TWOFNR as described in the text, multiplied by the spectroscopic factor.

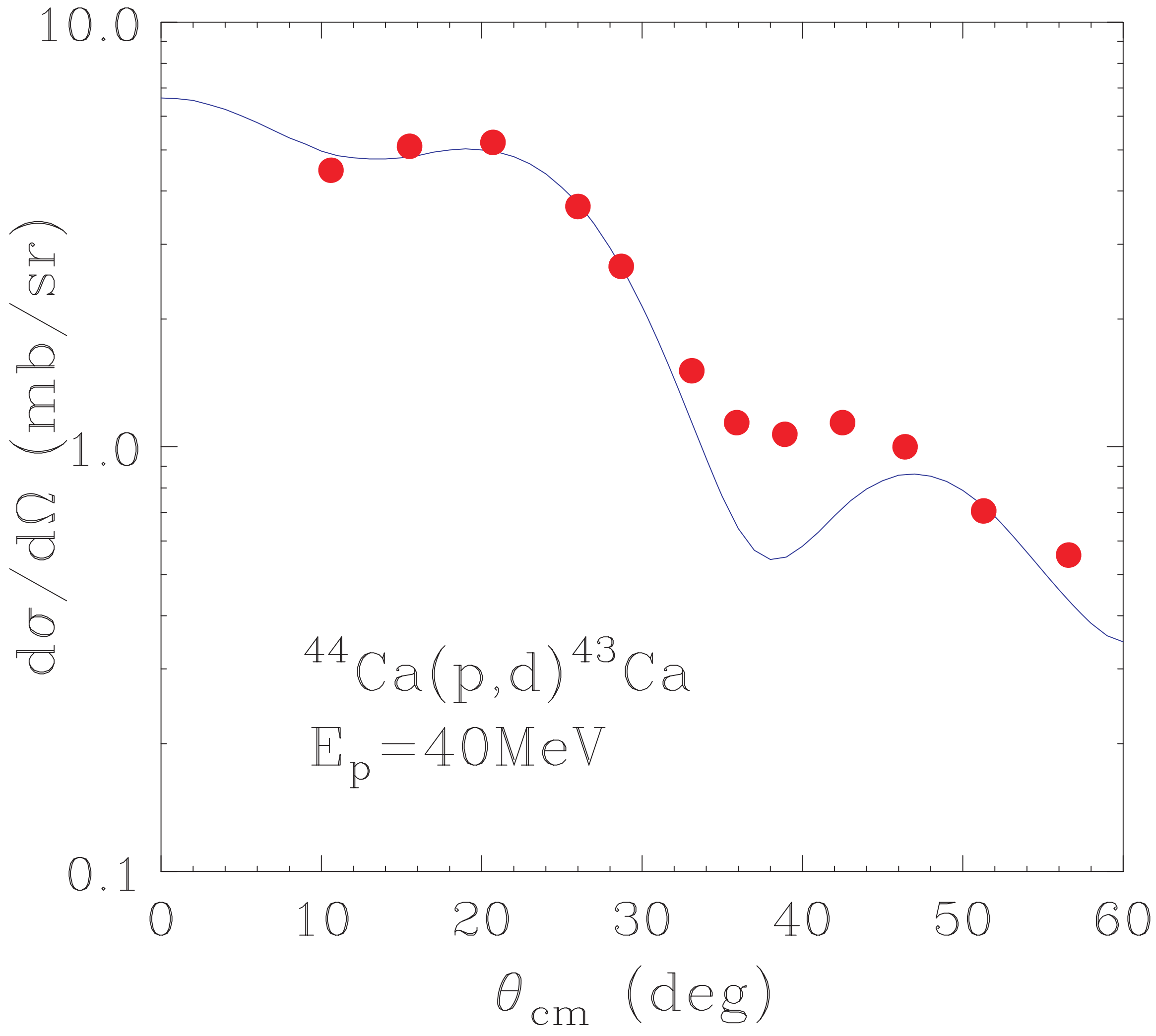

Figure 3: (Color online) Comparisons of the angular distributions of the proton measured in the $^{11}B(d,p)^{12}B$ reactions in three different experiments. Open circles, closed circles, open and closed diamonds represent data from refs. [21], [45] and [46] respectively.

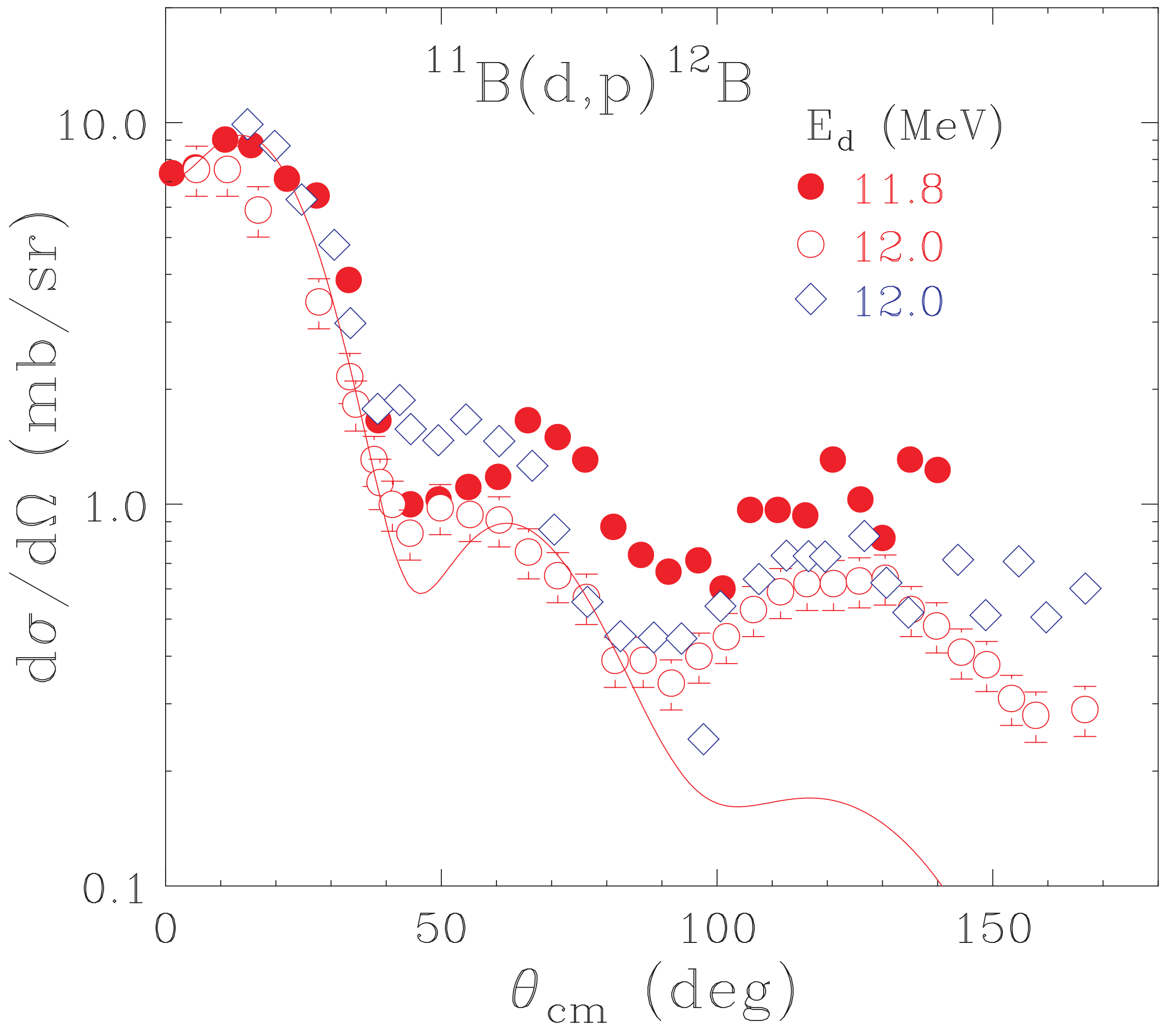

Figure 4: (Color online) Comparisons of the angular distributions of the proton measured in the $^{12}C(d,p)^{13}C$ reactions in four different experiments. Open circles and closed circles, open and closed diamonds represent data from refs. [21], [45], [59] and [30] respectively.

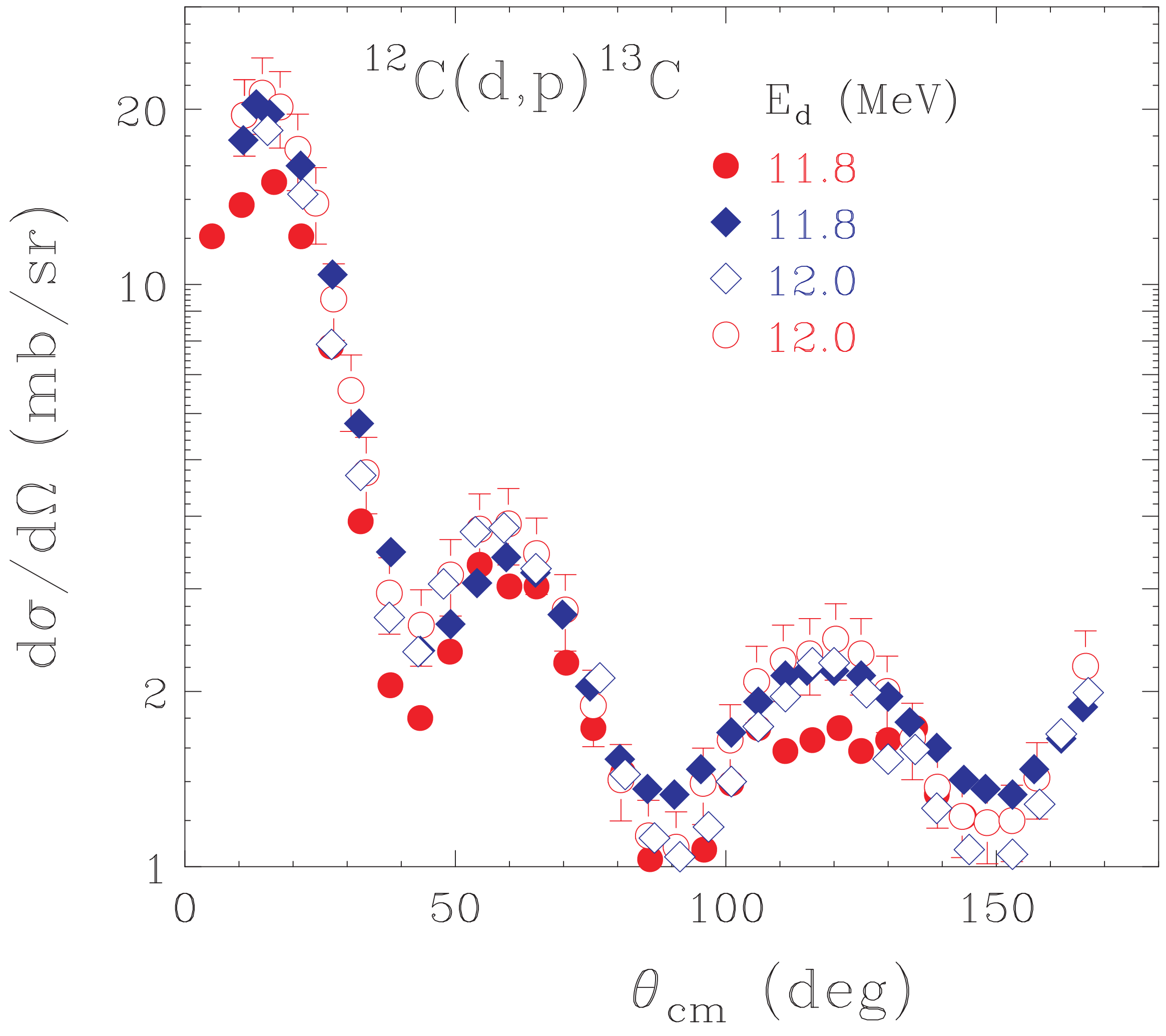

Figure 5: (Color online) Comparison of spectroscopic factors obtained from Ref. [181] (open circles) and from other measurements (closed circles). The increase of spectroscopic factors observed at $E_d$<10 MeV has been observed before [21] and has been attributed to the resonance structures in the elastic scattering of the deuterons [244].

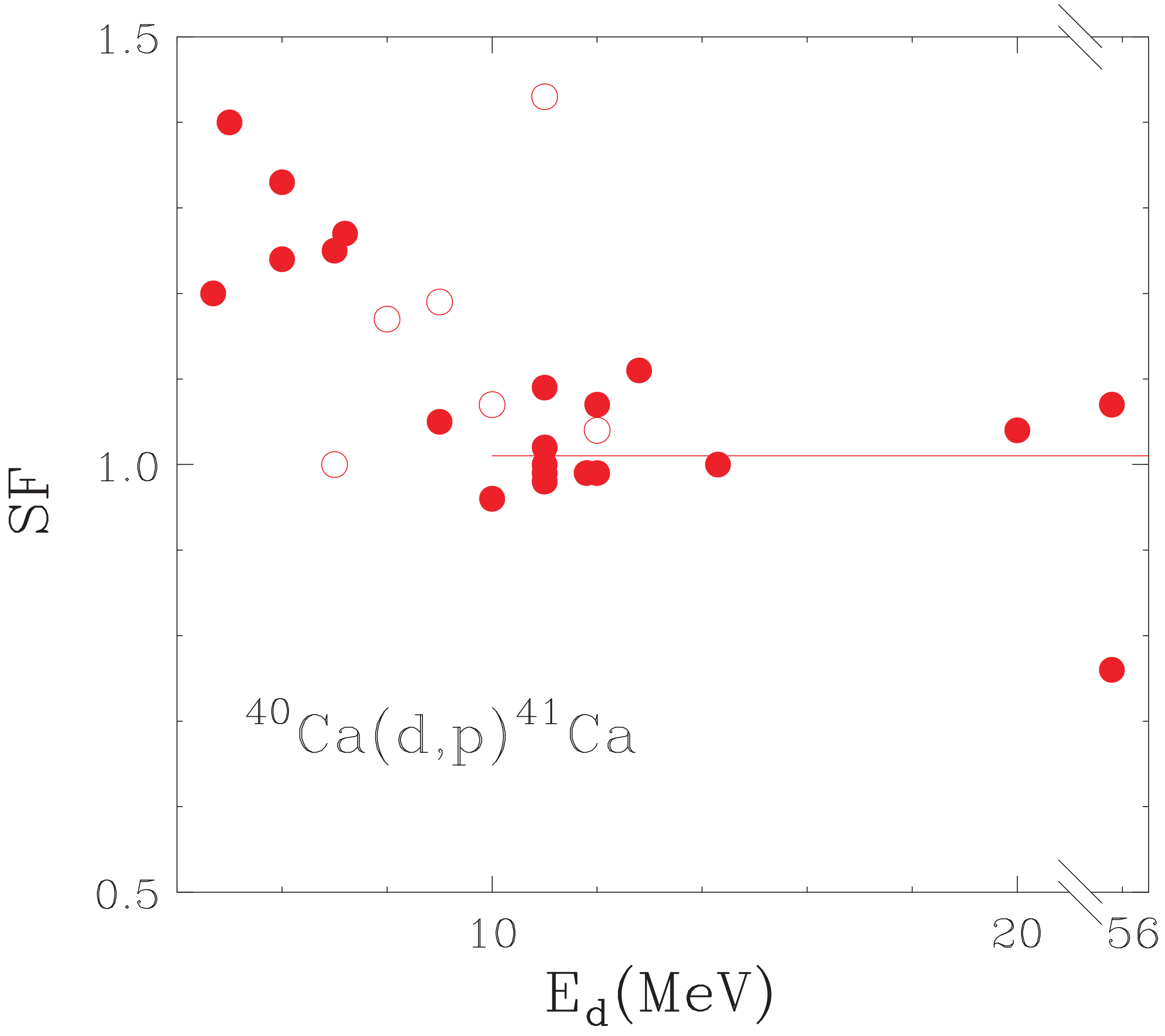

Figure 6: (Color online) Angular distributions for $^{40}Ca(d,p)^{41}Ca$ reactions for beam energy from 4.69 to 56 MeV. Each distribution is displaced by factors of 10 from adjacent distributions. The overall normalization factor is 10 for the 7.2 MeV data. References are listed in Table 1.

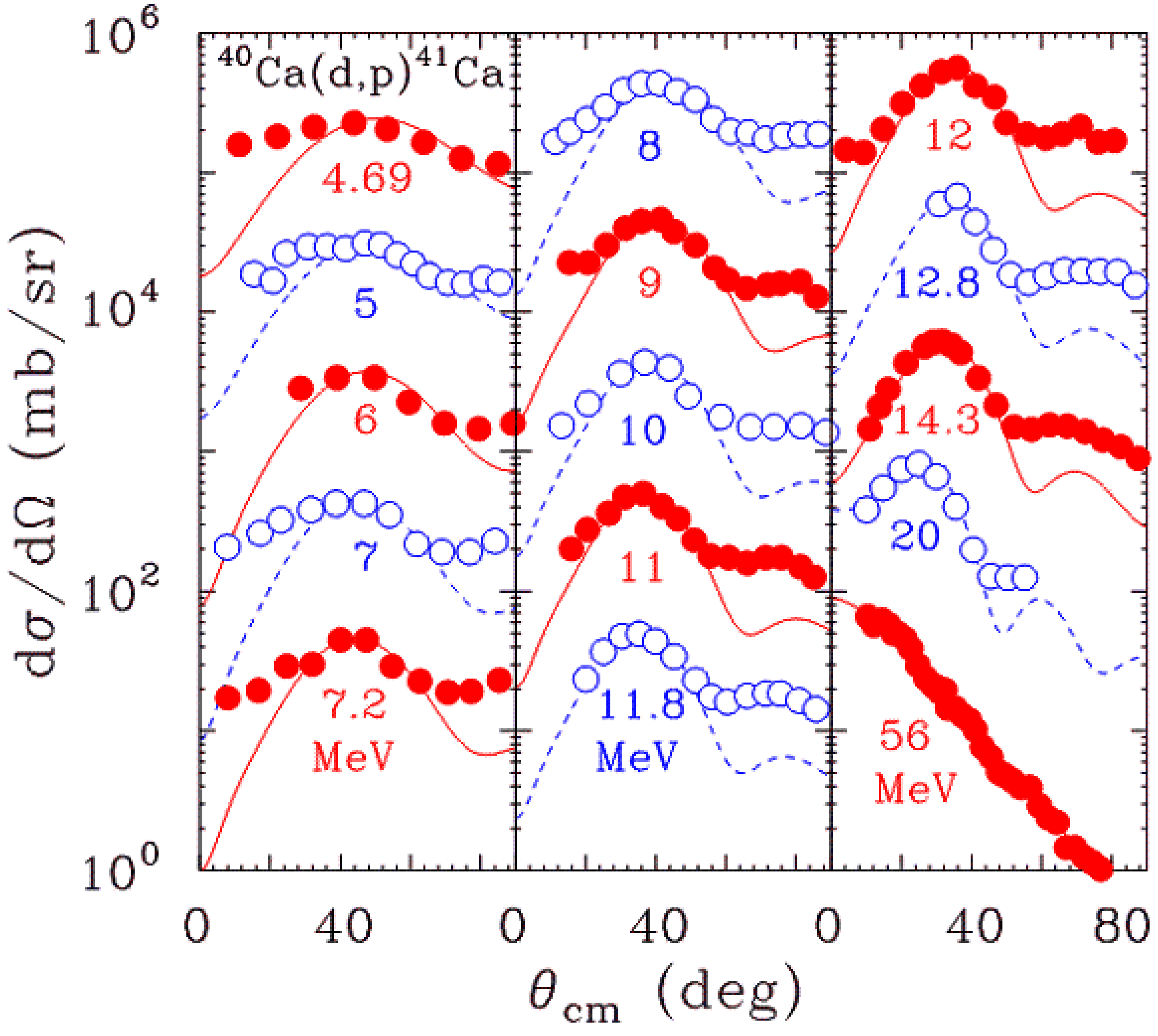

Figure 7: (Color online) Comparisons of the angular distributions of the deuteron measured in the $^{50}Cr(p,d)^{49}Cr$ reactions in two different experiments, closed and open circles are data from refs [223] and [224].

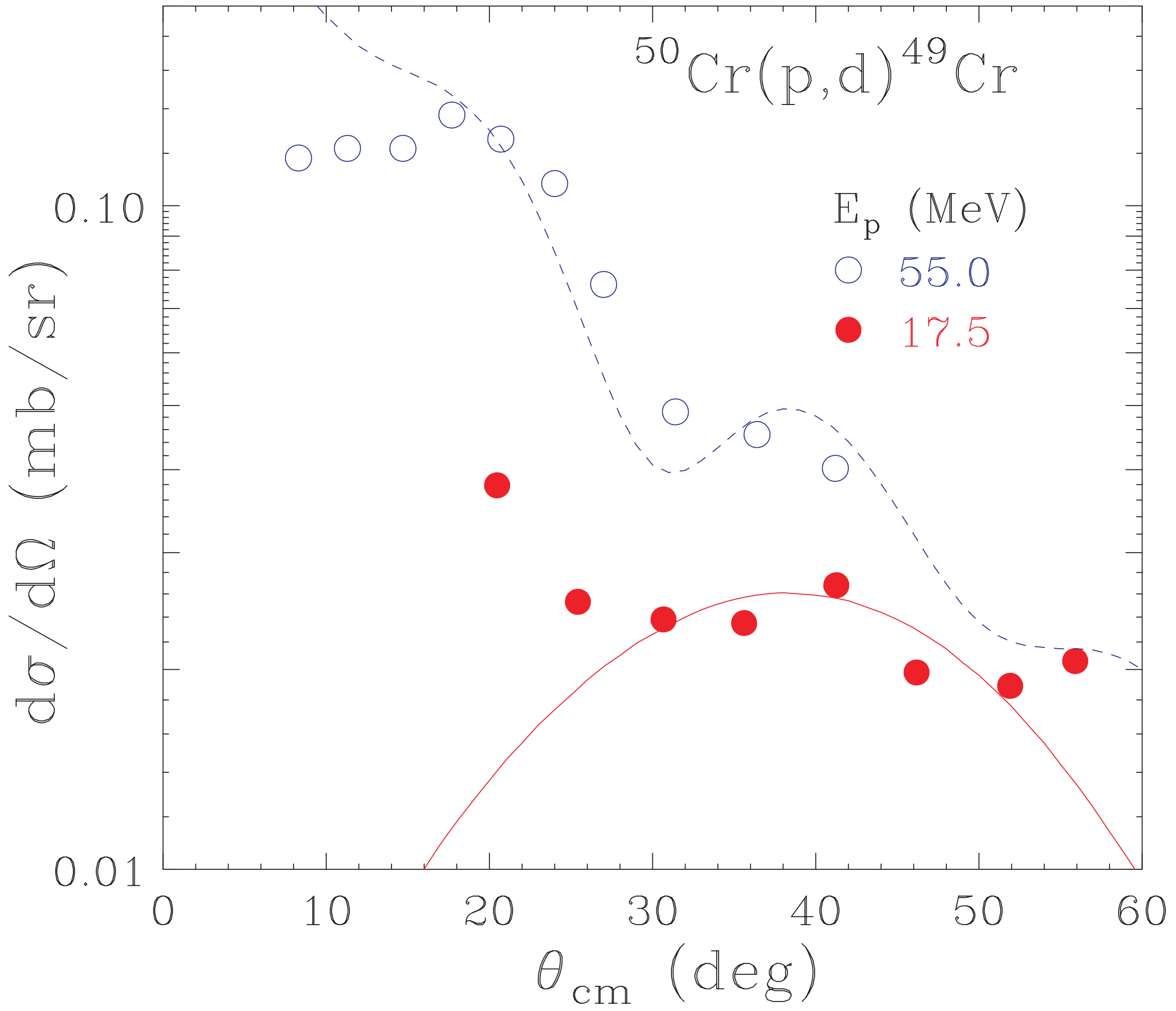

Figure 8: (Color online) Comparisons of spectroscopic factors obtained from (p,d) and (d,p) reactions as listed in Table II. The line indicates perfect agreement between the two values.

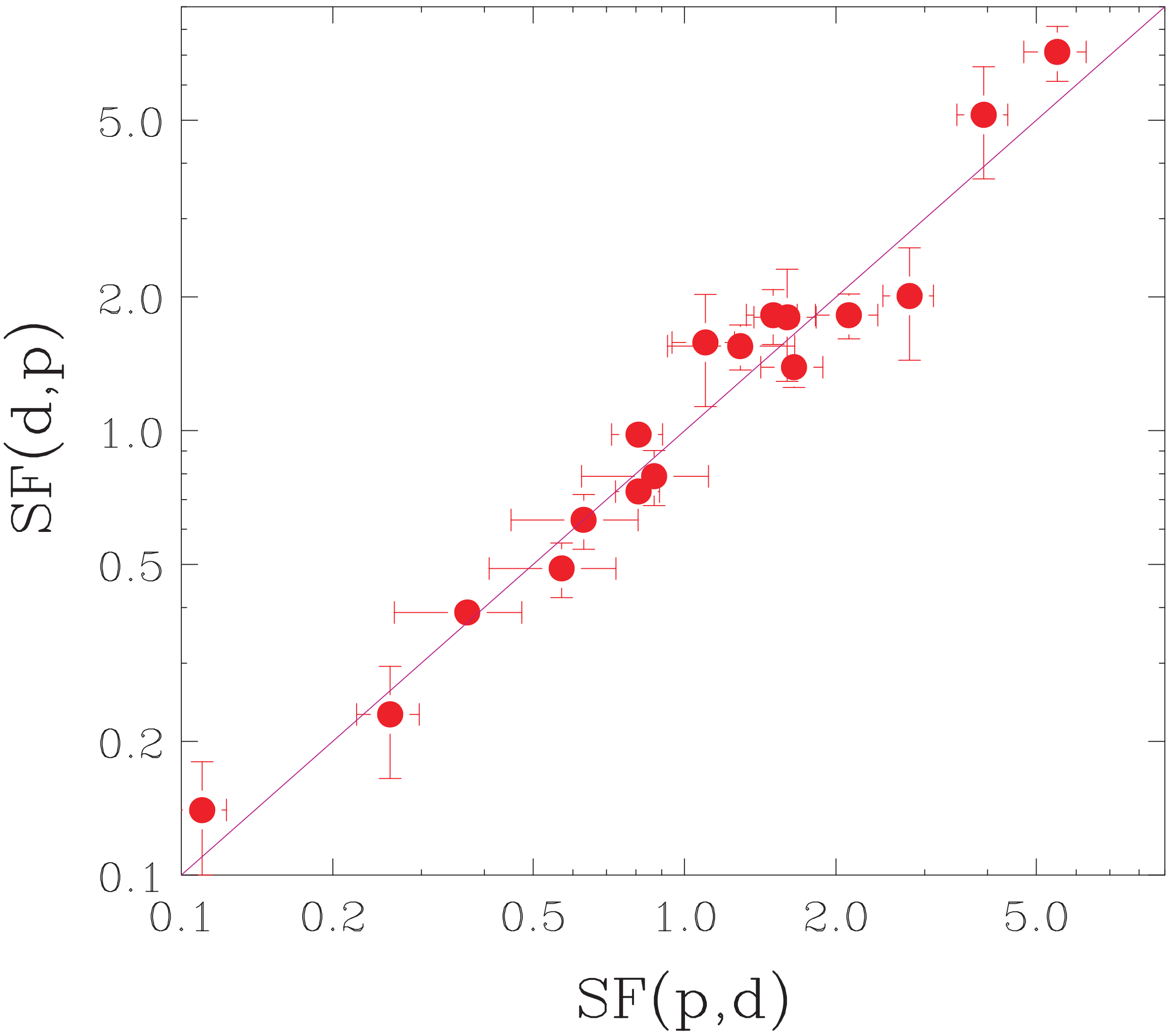

Figure 9: (Color online) Spectroscopic factors obtained for the Ca isotopes. The open stars represent individual measurements. The accompanying solid stars are the weighted averaged values with the associated uncertainties determined from the standard deviations or 20%/√N of the mean SF values whichever is larger.

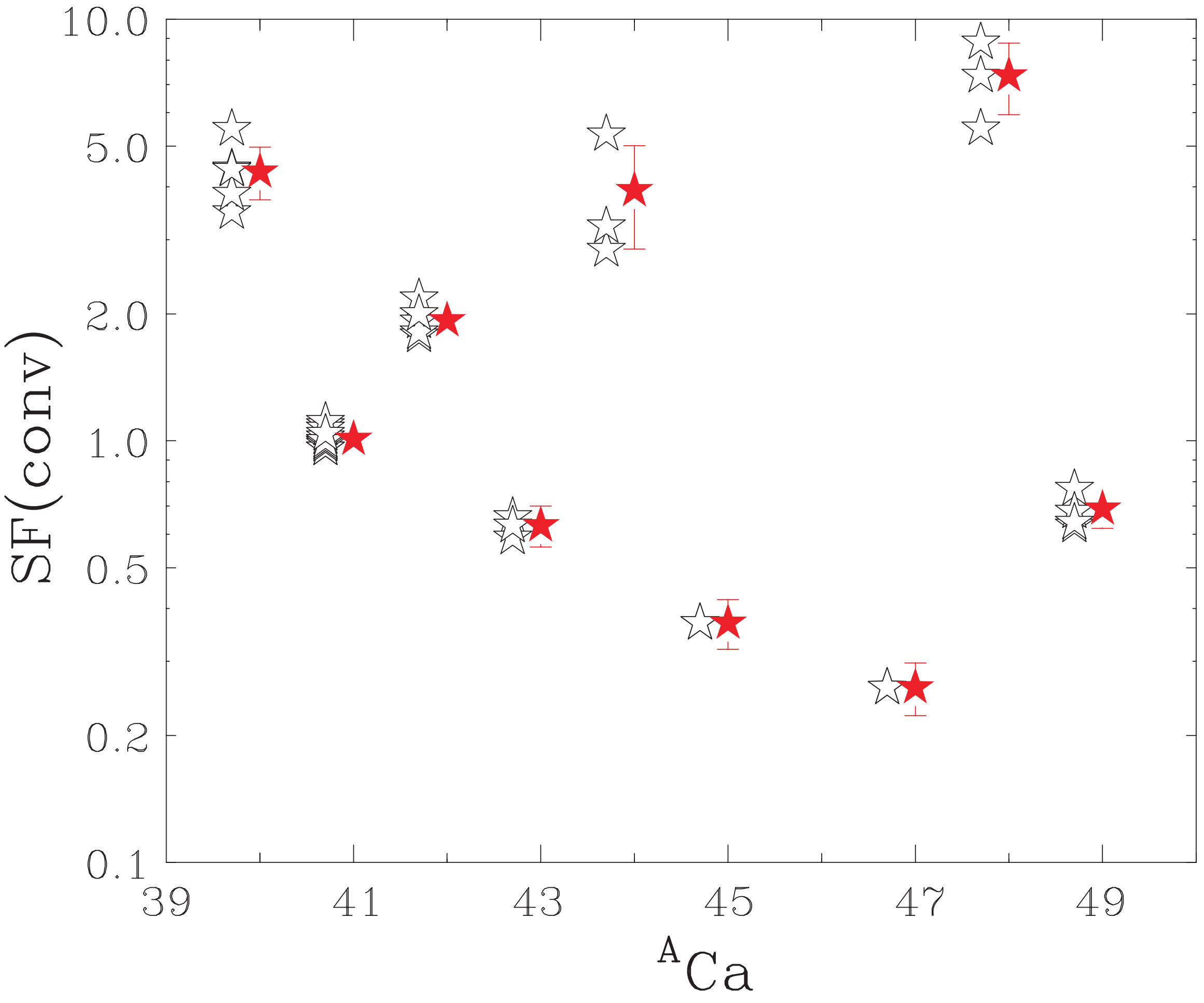

Figure 10: (Color online) Comparisons of spectroscopic factors obtained from this work SF(conv) and the compiled values of Endt [9]. All the values are listed in Table III. The line indicates perfect agreement between our values and Endt's compilation.

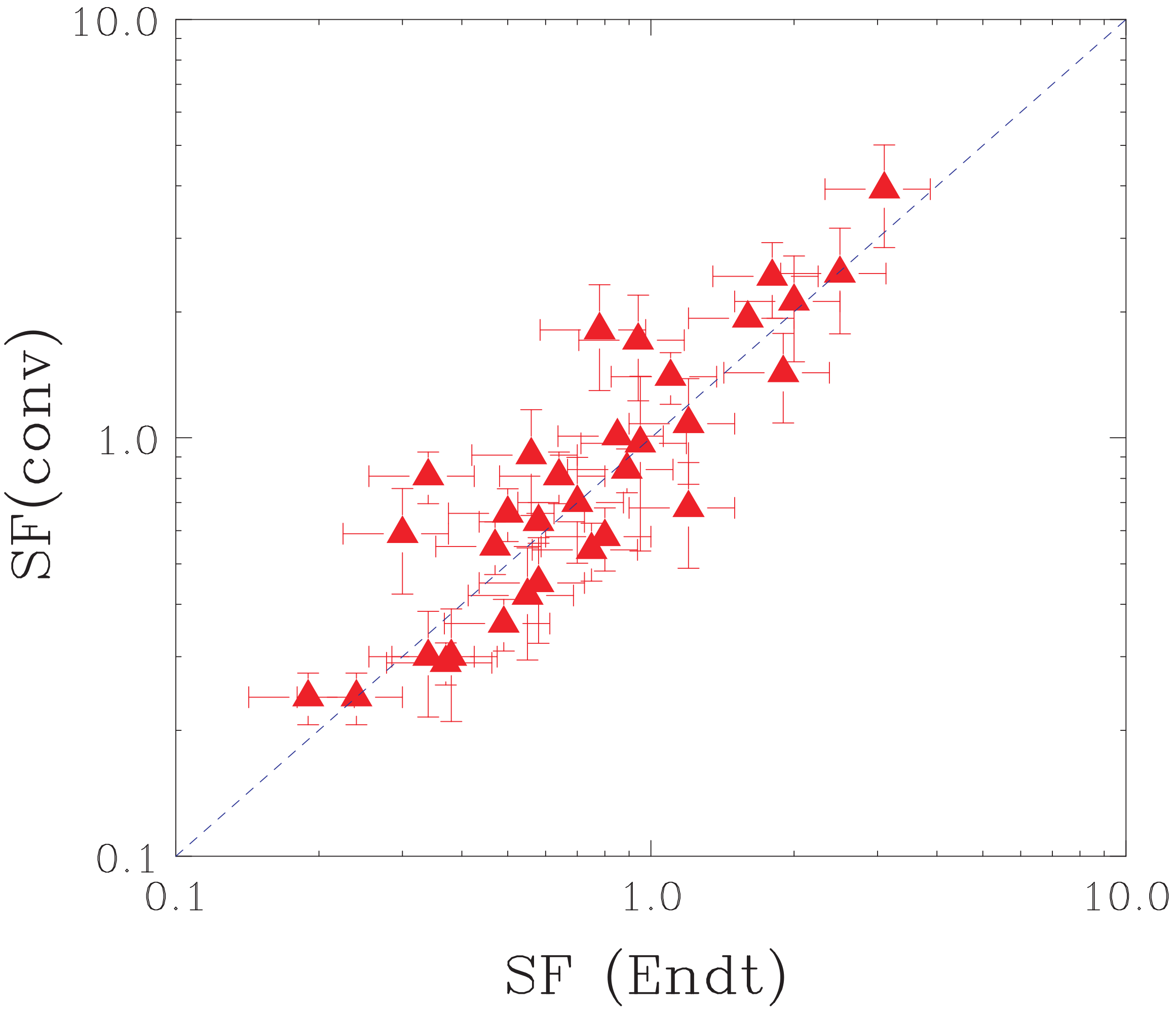

Figure 11: (Color online) Comparisons of three angular distributions of the deuteron measured in the $^{14}C(d,p)^{15}C$ reactions in three different experiments with incident deuteron energy of 14 MeV [74] (closed circles], 16 MeV [75] (closed squares] and 17 MeV [71] (open circles). The curve is the predicted angular distributions from the code TWOFNR as described in the text, multiplied by the spectroscopic factor of 1.1 which fits the data of ref. [71], the only set of data with measurements at angles more forward than 15°.

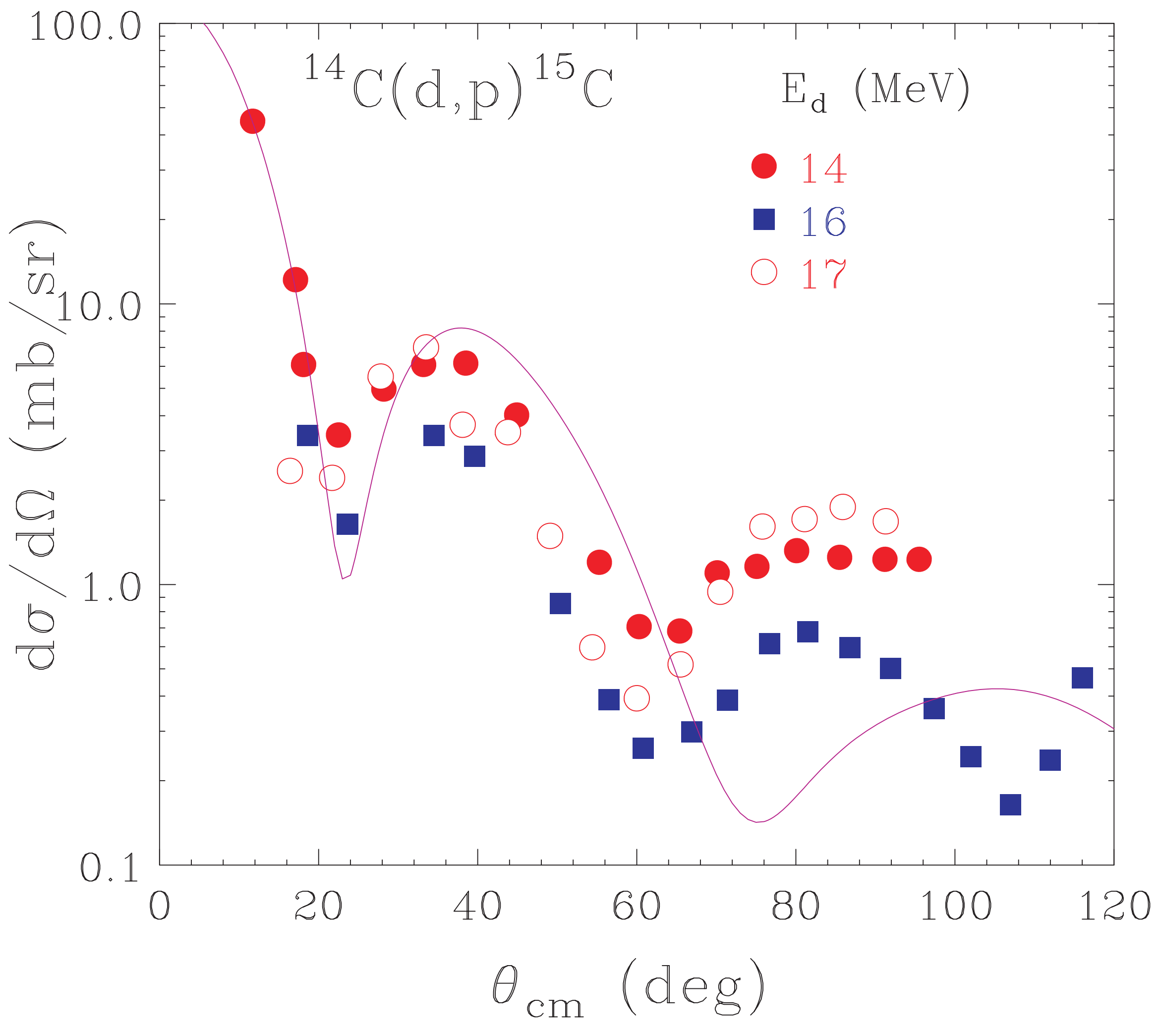

Figure 12: (Color online) Ratios of the SF values from experiment divided by the SF values predicted by the large basis shell model as a function of the neutron separation energy ($S_n$). Open and closed symbols denote elements with odd and even Z respectively.

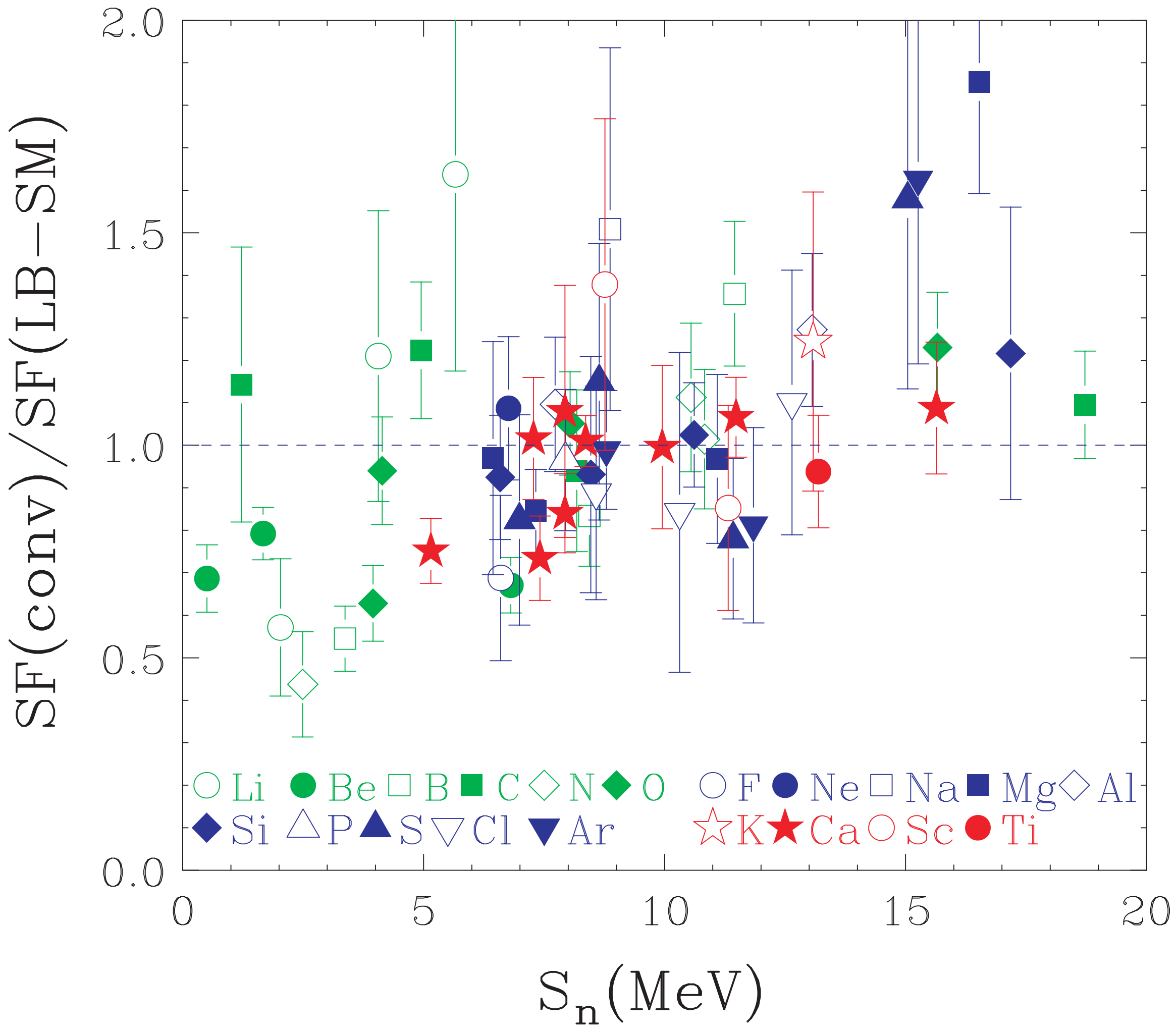